\documentclass[twocolumn,preprintnumbers,amsmath,amssymb,superscriptaddress,pre,showpacs]{revtex4}
\usepackage[english]{babel}
\usepackage[dvips]{graphicx}
\usepackage{dcolumn}
\usepackage{bm}
\usepackage[dvips]{epsfig}

\usepackage{color,ulem}

\newcommand{\gp}{\dot{\gamma}}

\begin{document}

\title{Timescales in creep and yielding of attractive gels}

 \author{Vincent Grenard}
 \affiliation{Universit\'e de Lyon, Laboratoire de Physique, \'Ecole Normale Sup\'erieure de Lyon, \\
 CNRS UMR 5672, 46 All\'ee d'Italie, 69364 Lyon cedex 07, France.}
\author{Thibaut Divoux}
\affiliation{Centre de Recherche Paul Pascal, CNRS UPR 8641, 115 Avenue Schweitzer 33600 Pessac, France.}
 \author{Nicolas Taberlet}
 \affiliation{Universit\'e de Lyon, Laboratoire de Physique, \'Ecole Normale Sup\'erieure de Lyon, \\
 CNRS UMR 5672, 46 All\'ee d'Italie, 69364 Lyon cedex 07, France.}
 \author{S\'ebastien Manneville}
 \affiliation{Universit\'e de Lyon, Laboratoire de Physique, \'Ecole Normale Sup\'erieure de Lyon, \\
 CNRS UMR 5672, 46 All\'ee d'Italie, 69364 Lyon cedex 07, France.}
 \affiliation{Institut Universitaire de France.}


\begin{abstract}
   
The stress-induced yielding scenario of colloidal gels is investigated under rough boundary conditions by means of rheometry coupled to local velocity measurements. Under an applied shear stress $\sigma$, the fluidization of gels made of attractive carbon black particles {dispersed} in a mineral oil is shown to involve a previously unreported shear rate response $\gp(t)$ characterized by two well-defined and separated timescales $\tau_c$ and $\tau_f$. First $\dot\gamma$ decreases as a weak power law strongly reminiscent of the primary creep observed in numerous crystalline and amorphous solids, coined the ``Andrade creep.'' We show that the bulk deformation remains homogeneous at the micron scale, which demonstrates that if plastic events take place or if any shear transformation zone exists, such phenomena occur at a smaller scale. As a key result of this paper, the duration $\tau_c$ of this creep regime decreases as a power law of the viscous stress, defined as the difference between the applied stress and the yield stress $\sigma_c$, i.e. $\tau_c\sim(\sigma-\sigma_c)^{-\beta}$, with $\beta=2$--3 depending on the gel concentration. The end of this first regime is marked by a jump of the shear rate by several orders of magnitude, while the gel slowly slides as a solid block experiencing strong wall slip at both walls, despite rough boundary conditions. Finally, a second sudden increase of the shear rate is concomitant to the full fluidization of the material which ends up being homogeneously sheared. The corresponding fluidization time $\tau_f$ robustly follows an exponential decay with the applied shear stress, i.e. $\tau_f = \tau_0 \exp(-\sigma/\sigma_0)$, as already reported for smooth boundary conditions. Varying the gel concentration $C$ in a systematic fashion shows that the parameter $\sigma_0$ and the yield stress $\sigma_c$ exhibit similar power-law dependences with $C$. Finally, we highlight a few features that are common to attractive colloidal gels and to solid materials by discussing our results in the framework of theoretical approaches of solid rupture (kinetic, fiber bundle, and transient network models).

\end{abstract}
\maketitle

\section{Introduction}

Yield stress fluids are ubiquitous in our everyday life and encompass a wide range of systems. From dry granular media \cite{Forterre:2008,Schall:2010}, slurries \cite{Stickel:2005}, and foams \cite{Hohler:2005,Katgert:2009} to (amorphous or crystallised) hard colloidal assemblies \cite{Chen:1990,Chen:1994b} and concentrated suspensions of soft colloidal particles such as emulsions, microgels \cite{Bonnecaze:2010}, etc. Despite a huge diversity of compositions and microstructures, the mechanical behavior of these materials is dominated by a critical shear stress $\sigma_c$, traditionally named the {\it yield stress} \cite{Moller:2009a}. Below $\sigma_c$, these materials all display a solid-like mechanical behavior which can be reversible as for a standard elastic solid, but most of the time involves aging phenomena \cite{Cloitre:2000,Ovarlez:2007,Negi:2010} and/or slow relaxation processes \cite{Ramos:2001b,Cipelletti:2005} depending on the particle elasticity, the packing fraction, and the nature of the interparticle interactions. For stresses above $\sigma_c$, the microstructure is fully reorganized and the yield stress fluid subsequently flows like a liquid while the shear rate reaches a steady-state value. Despite this apparently simple distinction between flowing and non-flowing states, the yielding transition still raises many open questions such as the influence of confinement \cite{Goyon:2008}, of boundary conditions \cite{Gibaud:2008}, or of previous flow history \cite{Ovarlez:2013}.

Among the above listed materials, colloids are a versatile model system which individual elastic properties and inter-particle interactions can be finely tuned \cite{Bonnecaze:2010,Royall:2013}. For packing fractions above $\phi_g \simeq 0.58$ colloidal hard spheres, characterized by repulsive interactions, display solid-like features at rest and behave as yield stress materials \cite{Sciortino:2005}. However, in the presence of attractive interactions, particles tend to stick to each other and an attractive colloidal dispersion behaves as a yield stress material from dense packing fractions (colloidal glasses) down to extremely low packing fractions (colloidal gels). The yielding scenario of colloidal assemblies for both repulsive and attractive interactions has been studied at length over the past fifteen years through different rheological approaches. First, yielding has been meticulously explored under {\it imposed shear rate} for hard spheres \cite{Derec:2003,Pham:2006,Zausch:2008,Zausch:2009,Koumakis:2012a,Amann:2013} and dense assemblies of soft spheres \cite{Carrier:2009,Divoux:2011a,Koumakis:2012b}, as well as for attractive glasses \cite{Pham:2008,Koumakis:2011} and gels \cite{Uhlherr:2005,Koumakis:2011,Mohraz:2005,Santos:2013}. For dense repulsive systems and low density gels, the yielding is a single step process involving a {\it stress overshoot} which amplitude increases for increasing shear rates \cite{Divoux:2011a,Koumakis:2011,Mohraz:2005,Amann:2013}. Based on confocal microscopy experiments and molecular dynamics simulations, it was recently shown that individual colloids in hard-sphere-like systems experience transient superdiffusive motions as they are being pushed out of their cage by shear \cite{Zausch:2008,Zausch:2009,Koumakis:2012a}. On the other hand sufficiently dense attractive systems exhibit a two-step yielding dynamics that involve two different characteristic strains which have been interpreted respectively as intercluster bond breaking followed by the breaking of clusters into smaller constituents \cite{Koumakis:2011}. 

This major difference in the yielding of attractive and repulsive non-ergodic systems has been confirmed through strain \cite{Petekidis:2002,Pham:2006,Smith:2007,Renou:2010,Datta:2011,Koumakis:2012b,Shao:2013} or stress \cite{Dimitriou:2013} controlled {\it Large Amplitude Oscillatory Shear} (LAOS) (see Ref.~ \cite{Hyun:2011} for a recent enlightening review). Starting from low oscillation amplitudes, the elastic component $G'$ and the viscous component $G''$ of the shear modulus remain fairly constant with $G'\gg G''$ for any solidlike colloidal assembly over the usual range of accessible frequencies ($f=10^{-3}$--10~Hz). In the case of dense hard sphere-like systems, upon increasing the strain amplitude, $G'$ is observed to decrease monotonically as a power law and intersects $G''$, which exhibits a bell-shaped curve, defining a single yield strain associated to sample fluidization \cite{Pham:2006}. In the case of attractive colloidal glasses, $G'$ departs sooner from the linear regime and does not decrease monotonically but exhibits a second shoulder at a higher strains, which defines a second yield strain where the local topology of a particle, bounded to its nearest neighbours, is supposedly destroyed \cite{Pham:2006}. Thus, LAOS experiments also support the picture of a single vs a two-step yielding scenario in repulsive and attractive colloidal glasses respectively. Oscillatory experiments performed at constant strain-rate rather than constant frequency, and coined {\it Strain Rate Frequency Superposition}, nicely supplement LAOS experiments since they have allowed to probe the universality of the structural relaxation in dense colloidal assemblies and to put forward a shear-rate dependent timescale associated to the sample yielding \cite{Wyss:2007,Datta:2011}.  

Finally, the yielding scenario has been studied through step-stress experiments starting from rest. Previous work on colloids include a vast series of papers on hard sphere glasses \cite{Petekidis:2003,Petekidis:2004,Siebenburger:2012}, soft sphere glasses \cite{Coussot:2006,Caton:2008,Paulin:1997,Divoux:2011a}, attractive glasses \cite{Laurati:2011} and colloidal gels \cite{Uhlherr:2005,Coussot:2006,Gopalakrishnan:2007,Gibaud:2010,Sprakel:2011,Lindstrom:2012}. For stresses applied below the yield stress the material experiences a creep flow and eventually stops flowing, while the shear rate decreases as a weak power law of time \cite{Paulin:1997,Coussot:2006,Divoux:2011a,Siebenburger:2012}. Above the yield stress, the material first displays a creep regime during which the shear rate may either be constant or decrease as a weak power law, followed by a brutal fluidization associated to an increase of the shear rate by several orders of magnitude before the shear rate reaches steady state \cite{Coussot:2006,Caton:2008,Gibaud:2010,Divoux:2011a,Siebenburger:2012}. The duration of the creep regime prior to abrupt yielding has been reported to decrease for increasing applied shear stress \cite{Caton:2008,Gibaud:2010,Divoux:2011a,Sprakel:2011}. For applied stresses close to the yield stress, the creep regime may become extremely long (up to several hours) which is often referred to as ``delayed yielding" or ``time-dependent yielding'' in the literature \cite{Uhlherr:2005,Lindstrom:2012}. Now, contrary to what has been discussed above for shear startup and LAOS experiments, the yielding scenario under applied stress appears to be surprisingly similar for both attractive and repulsive colloidal assemblies. For both types of materials, the yielding appears to be a single-step process well characterized by a single timescale (which decreases for increasing applied shear stress) and a shear rate that exhibits a temporal evolution with a characteristic ``S" shape.

To summarize this overview of the current state of knowledge, it is now well established based on the various protocols listed above that a yield stress departs a solid state from a liquid one and that yielding involves complex system-dependent dynamical processes that reflect in the behaviour of rheological observables such as the shear stress $\sigma$, the shear rate $\dot{\gamma}$, or viscoelastic moduli $G'$ and $G''$. However, in spite of a few experimental and numerical approaches at the scale of individual colloids already mentioned above  \cite{Zausch:2008,Zausch:2009,Koumakis:2012a}, the question remains whether the various timescales or characteristic strains involved in yielding have a mesoscopic signature on the local strain field, e.g. through collective behaviour, fracture planes, or shear localization. In the past few years the transient regime leading from rest to flow has been shown to involve either heterogeneous \cite{Divoux:2010,Divoux:2011a,Rogers:2008} or homogeneous \cite{Besseling:2010} flows, themselves leading to either heterogeneous or homogeneous flows in steady state \cite{Ovarlez:2009,Ovarlez:2013b}. While several mechanisms have been proposed and tested to account for steady-state flow heterogeneities, including competition between aging and shear rejuvenation \cite{Moller:2008} or flow-concentration coupling \cite{Besseling:2010}, the reason why soft glassy materials shall exhibit homogeneous rather than heterogeneous velocity profiles during {\it transient} flows remains a burning issue \cite{Moorcroft:2013}. In particular the role of the interparticle interactions \cite{Chaudhuri:2012} and the subtle interplay between the flow and the boundary conditions \cite{Buscall:1993,Gibaud:2008,Gibaud:2009,Buscall:2010,Seth:2012} remain to be fully assessed.

The aim of the present paper is to investigate thoroughly the yielding dynamics of attractive colloidal gels made of carbon black particles under steady shear stress. The stress-induced yielding of such attractive gels has already been studied but only under smooth boundary conditions \cite{Gibaud:2010,Sprakel:2011} or with no access to the local fluidization scenario \cite{Gopalakrishnan:2007}. Here we demonstrate through rheological measurements coupled to time-resolved velocimetry that stress-induced yielding of this attractive colloidal assembly involves a two-timescale process in the presence of rough boundary conditions. After an initial creep regime, the gel first fails at the moving wall with a timescale $\tau_c$ that depends on the applied shear stress and decreases as a power law of the viscous stress, defined as the difference between the stress and the yield stress,  i.e. $\tau_c\sim(\sigma-\sigma_c)^{-\beta}$, with $\beta=2$--3 depending on the gel concentration. Second, the attractive gel slowly turns from a sliding solid block into a fully liquid state through a fluidization process that  involves strongly heterogeneous flows and characterized by another timescale $\tau_f$ that decays exponentially with the applied shear stress, i.e. $\tau_f = \tau_0 \exp(-\sigma/\sigma_0)$. These two timescales $\tau_c$ and $\tau_f$ are measured for various gel concentrations $C$. We show that the characteristic stresses $\sigma_c$ and $\sigma_0$ depend on $C$ as power laws with exponents comparable to those found for standard rheological quantities. This two-step yielding scenario under rough boundary conditions is significantly different from that observed in smooth geometries where only one timescale $\tau_f$ is reported and from that observed for dense repulsive microgels \cite{Divoux:2011a}. Finally our results are discussed in light of the experiments mentioned in the introduction and of models for solid rupture.

\section{Materials and methods}

\subsection{Experimental setup}

Our experimental setup for performing simultaneous rheological and local velocity measurements has been described at length in Ref.~ \cite{Manneville:2004a}. It consists in coupling a standard stress-controlled rheometer (Anton Paar MCR301 in the present work) to an ultrasonic velocimetry technique that allows one to monitor the azimuthal velocity profile within the 1-mm gap of a Taylor-Couette cell with a spatial resolution of about 40~$\mu$m. Ultrasonic velocimetry relies on the cross-correlation of successive pressure signals backscattered by the sample. Acoustic contrast arises either naturally from the material microstructure or artificially through seeding with small contrast agents such as glass or polystyrene microspheres. The reader is referred to Ref.~ \cite{Manneville:2004a} for full details on ultrasonic data analysis.

In the following, we shall note respectively $\sigma$ and $\gp$ the shear stress and shear rate imposed or recorded by the rheometer. It is important to keep in mind that these are ``engineering'' or ``global'' quantities, in the sense that they result from torque and velocity measurements on the inner rotating cylinder. In particular, in the presence of wall slip, $\gp$ only represents an apparent shear rate which may strongly differ from the actual shear rate in the bulk material. \cite{Barnes:1995}

In our previous work on carbon black gels, only ``smooth'' boundary conditions were used  \cite{Gibaud:2010}. Here, in order to focus on the effect of boundary conditions on yielding, we shall rather use sand-blasted Plexiglas for both the fixed outer cylinder and the inner rotating bob of our Taylor-Couette device. Sand-blasting induces a typical surface roughness of 1~$\mu$m which is large enough to scatter ultrasound significantly. In order to avoid artifacts due to ultrasonic waves scattered off the rough stator, a specific procedure has been implemented as detailed in the supplementary material (see Supplemental Fig.~1). 

\subsection{Rheological characteristics of the samples}
\label{s.charac}

\begin{figure}[!t]\tt
\centering
\includegraphics[width=\columnwidth]{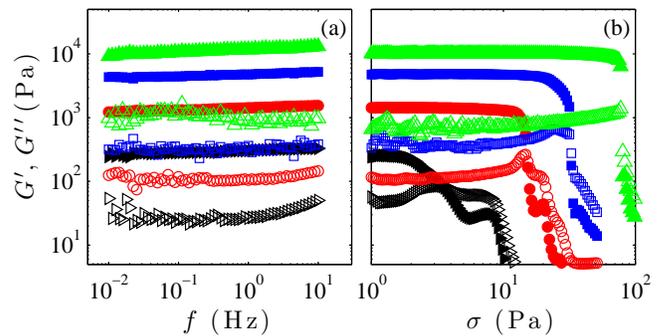}
\caption{Viscoelastic moduli $G'$ (filled symbols) and $G''$ (open symbols) of carbon black gels at $C=4$ ($\triangleright$), 6 ($\circ$), 8 ($\square$), and 10\%~w/w ($\triangle$) in a rough Couette geometry. (a)~Frequency sweeps at a fixed stress amplitude $\sigma=2$~Pa. The waiting time per point is 6 oscillation periods. (b)~Stress sweeps at a fixed frequency $f=1$~Hz with a waiting time of 5~s per point.}
\label{f.rheol}
\end{figure}

\begin{figure}[!t]\tt
\centering
\includegraphics[width=0.9\columnwidth]{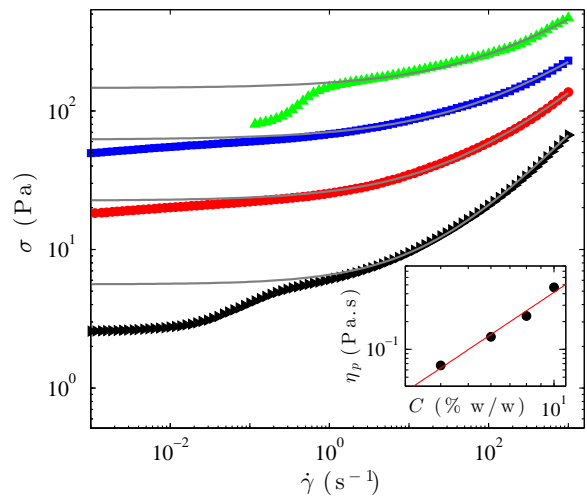}
\caption{Flow curves $\sigma$ vs $\gp$ of carbon black gels at $C=4$, 6, 8, and 10\%~w/w from bottom to top, in a rough Couette geometry. The shear rate is swept down from $\gp=10^3$~s$^{-1}$ with a logarithmic spacing of 15 points per decade and a waiting time of 1~s per point. Solid lines are fits using a Herschel-Bulkley law, $\sigma=\sigma_{y2}+A\gp^n$, for $\gp>1$~s$^{-1}$. For $C=10\%$ w/w, shear rates smaller than 0.1~s$^{-1}$ could not be reached with such a short waiting time per point. Inset: viscosity $\eta_p$ measured at the highest shear rate $\gp_p=10^3$~s$^{-1}$ as a function of the gel concentration $C$. The red line is the best power-law fit which yields an exponent of 2.1.}
\label{f.nlrheol}
\end{figure}

Carbon black (CB) refers to colloidal soot particles produced from the incomplete combustion of fossil fuels. Made of permanently fused ``primary" particles of diameter 20--40~nm \cite{Samson:1987}, these fractal soot particles, of typical size 0.2--0.5 $\mu$m, tend to form reversible and weakly linked agglomerates when dispersed in a liquid hydrocarbon \cite{Waarden:1950}. Indeed CB particles interact through a short-range attractive potential, whose depth is estimated at about 30~$k_BT$ \cite{Trappe:2007}. Small volume fractions are sufficient to turn the dispersion into an interconnected network, {hence into a colloidal gel} \cite{Trappe:2000,Trappe:2001}. Such {dispersions} are involved in a wide variety of industrial applications including paints, coatings, rubbers and tires \cite{Donnet:1993}. On a more fundamental side, CB colloidal gels, which properties may be tuned by the addition of dispersant \cite{Trappe:2000,Won:2005}, are model systems for studying both the yielding transition \cite{Gibaud:2010,Sprakel:2011} and the subsequent shear-induced structuration in confined geometries \cite{Osuji:2008a,Grenard:2011}, since they are neither subject to shear degradation nor to evaporation and thus allow long lasting and reproducible measurements.

Here, CB gels are prepared in the absence of any dispersant by mixing CB particles (Cabot Vulcan XC72R of density 1.8) in a light mineral oil (from Sigma, density 0.838, viscosity 20~mPa.s) as described in Ref.~ \cite{Gibaud:2010} at weight concentrations $C$ ranging from 4 to 10\%~w/w, which roughly correspond to effective volume fractions from 0.1 to 0.3 \cite{Trappe:2007}. Furthermore, in order to ensure ultrasonic scattering, the gels are seeded with 1\%~w/w hollow glass microspheres of mean diameter 6mm (Sphericel, Potters) and density 1.1~kg.m$^{-3}$.

All the experiments reported in the present paper are performed at a temperature of 25$^\circ$C which is held fixed up to $\pm 0.1^\circ$C thanks to a water circulation around the Couette cell that also ensures the acoustic coupling between the ultrasonic probe and the cell. CB dispersions are systematically presheared at a high shear rate (typically $\gp_p=10^3$~s$^{-1}$ for 20~s). As discussed by Osuji {\it et al.}  \cite{Osuji:2008b,Negi:2009a}, such preshearing allows the sample to reach a shear-thickened state due to the breakup of locally dense clusters into sparser structures, which enhances viscous dissipation. For concentrations larger than 4\%~w/w, shear-thickening always sets in below 1000~s$^{-1}$ so that our preshearing rate $\gp_p$ is into the shear-thickening regime for all concentrations  \cite{Osuji:2008b,Kawaguchi:2001}. Upon flow cessation, very fast gelation of the system is observed over less than 1~s leading to a solidlike behaviour.

\begin{table}[b]
\small
  \caption{\small Rheological parameters of carbon black gels at different weight concentrations $C$ (see text for definitions). The last line gives the values of the exponents extracted from the power-law fits shown in Fig.~\ref{f.rheolC}.}
  \label{t.rheol}
  \begin{tabular*}{0.49\textwidth}{@{\extracolsep{\fill}}llllllll}
    \hline
     $C$ & $G'$ (Pa) &  $G''$ (Pa) & $\sigma_{y_1}$ (Pa) & $\sigma_{y_2}$ (Pa) & $n$ & $\eta_p$ (Pa.s)  \\
    \hline
    4\% &  310         & 30 & 4.8 & 5.6 & 0.60 & 0.067\\
    6\% &  1500	& 110  & 15.3 & 22.5 & 0.49 & 0.137\\
    8\% &   4900 	& 380 & 32.5 & 62 & 0.43 & 0.230\\
    10\% &  1.2$\cdot$10$^{4}$	&820 & 80 & 146 & 0.45 & 0.468\\
    \hline
exp. & 4.0 & 3.7 & 3.0 & 3.5 & n/a & 2.1\\
  \hline
  \end{tabular*}
\end{table}

\begin{figure}[!t]\tt
\centering
\includegraphics[width=0.9\columnwidth]{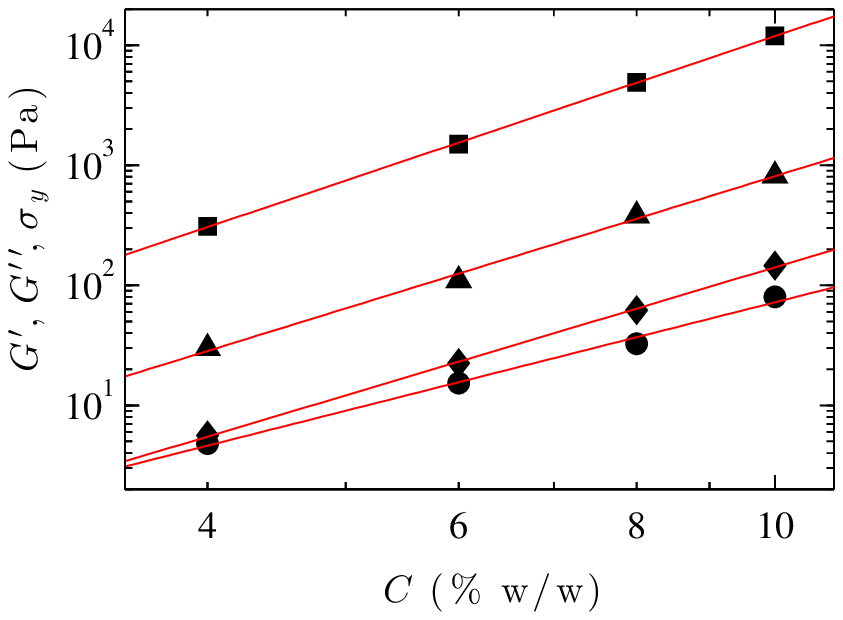}
\caption{Rheological parameters gathered in Table~\ref{t.rheol} and plotted against the concentration $C$. Storage modulus $G'$ ($\blacksquare$) and loss modulus $G''$ ($\blacktriangle$) measured at $f=1$~Hz and $\sigma=2$~Pa. Yield stress $\sigma_{y1}$  inferred from the crossing between $G'$ and $G''$ in an oscillatory stress sweep at $f=1$~Hz ($\bullet$). Yield stress $\sigma_{y2}$ ($\blacklozenge$) inferred from the Herschel-Bulkley fits of the flow curves shown in Fig.~\ref{f.nlrheol}. Red lines are the best power-law fits whose exponents are given in the last line of Table~\ref{t.rheol}.}
\label{f.rheolC}
\end{figure}

The rheological features of the resulting gels are captured in Figs.~\ref{f.rheol} and \ref{f.nlrheol}. Figure~\ref{f.rheol} shows oscillatory shear tests performed on the various gels involved in the present study. In the linear regime the evolutions of the viscoelastic moduli $G'$ and $G''$ with frequency [Fig.~\ref{f.rheol}(a)] are characteristic of soft solids with a storage modulus $G'$ that remains essentially constant and much larger than the loss modulus $G''$, which shows a small increase at high frequency  \cite{Kawaguchi:2001}. Stress sweeps at a given frequency [Fig.~\ref{f.rheol}(b)] allow one to get an estimate for the yield stress $\sigma_{y1}$ by looking at the stress amplitude at which $G'=G''$. Table~\ref{t.rheol} and Figure~\ref{f.rheolC} summarize the characteristic rheological features of our samples as a function of their weight concentration $C$. The various rheological parameters are seen to increase with $C$ as power laws of exponents ranging from 3 to 4. In particular the elastic modulus $G'\sim C^{\alpha}$ yields an exponent of 4.0 in line with previous measurements on shear-thickened CB gels prepared in similar low polarizable solvents such as tetradecane \cite{Osuji:2008b} ($\alpha = 3.5$) or  base stock oil \cite{Trappe:2000} ($\alpha = 4.1$). Also note that increasing the solvent molecular mass or increasing its polarizability decreases the degree of flocculation \cite{Waarden:1950} and leads to weaker structures \cite{Hiemenz:1965} which explains the smaller values of the exponents \cite{Khan:1993} also found in the literature, e.g. $\alpha=2.6$ for CB dispersed in silicone oil \cite{Kawaguchi:2001}. Interestingly, the exponent $4.0$ that we report here is also consistent with earlier data on gels of (i)~silica particles coated with octadecyl chains and dispersed in hexadecane \cite{Chen:1991} ($3.2 \leq \alpha \leq 7 $), or decalin \cite{Rueb:1997} ($4.4 \leq \alpha \leq 5.6$), (ii)~glycerol tristearate aggregates in olive oil \cite{Vreeker:1992} ($\alpha=4.1$), (iii)~bohemite alumina powders dispersed in water \cite{Shih:1990} ($\alpha=4.1$), and (iv)~aggregated polystyrene latex dispersions \cite{Buscall:1988,Rooij:1994} ($\alpha=4.6\pm 0.3$). These exponents have been linked to a gelation mechanism based on the aggregation of fractal clusters \cite{Buscall:1988} and a scaling approach has shown that well above the gelation threshold, the elastic properties are dominated by the fractal nature of the flocs that form the building blocks of the system \cite{Shih:1990}. In particular, since the yield strain defined as $\gamma_c=\sigma_{y_1}/G'$ decreases for increasing carbon black gel concentrations (see Table~1), we can expect from Ref.~ \cite{Shih:1990} that the gels studied in this paper are made of large and weak flocs, and that their yielding process will be dominated by bond breaking within each floc.

The observed scaling of the elastic modulus with concentration can be further interpreted in terms of the relationship $G'\sim C\sigma_p^{d_f/(3-d_f)}$ derived in Ref.~ \cite{Osuji:2008b} where $\sigma_p$ is the preshear stress at which the gel is prepared and $d_f$ the cluster fractal dimension. This relationship results from the scaling of the size $R_c$ of fractal clusters with $\sigma_p$, $R_c\sim\sigma_p^{1/(d_f-3)}$, and from $G'\sim CU/R_c^{d_f}$, where $U$ is the interaction energy between two colloids. In our case, CB gels are prepared at a given preshear rate $\gp_p=10^3$~s$^{-1}$ whatever the concentration $C$. Introducing the viscosity in the presheared state $\eta_p=\sigma_p/\gp_p$ and using $G'\sim C^\alpha$, one expects $\eta_p\sim C^{(\alpha-1)(3-d_f)/d_f}$. As shown in the inset of Fig.~\ref{f.nlrheol}, $\eta_p$ indeed follows a power-law with an {exponent $\alpha'=2.1$ (see also Table~\ref{t.rheol}). Assuming that $\alpha'=(\alpha-1)(3-d_f)/d_f$ and using $\alpha=4.0$, one gets $d_f=1.8\pm 0.1$ in full agreement with thermoporometry that reports $d_f=1.81$--1.85$\pm 0.05$ for the same CB particles dispersed in water and undecane \cite{Ehrburger-Dolle:1990}. This consistency} provides a nice confirmation of the scaling of the viscoelastic modulus with $C$ and $\sigma_p$ already found in Ref.~ \cite{Osuji:2008b}.

Flow curves $\sigma$ vs $\gp$ measured by rapidly sweeping down the shear rate from $\gp=10^3$~s$^{-1}$ are shown for concentrations $C=4$, 6, 8, and 10\%~w/w in Fig.~\ref{f.nlrheol}. For shear rates larger than 1~s$^{-1}$, the data are reasonably well fitted by Herschel-Bulkley behaviours, $\sigma=\sigma_{y2}+A\gp^n$. This provides estimates for the yield stress $\sigma_{y2}$ that are compared to $\sigma_{y1}$ in Table~\ref{t.rheol} and in Fig.~\ref{f.rheolC}. The exponent $n$, which ranges from 0.3 to 0.6, is typical of values reported in the literature for a wide variety of yield stress fluids such as concentrated emulsions  \cite{Salmon:2003a}, foams  \cite{Hohler:2005,Marze:2008,Katgert:2009,Ovarlez:2010}, and microgels  \cite{Roberts:2001,Cloitre:2003,Coussot:2009,Divoux:2012}. It shows a decrease with increasing concentrations, i.e. shear-thinning is more pronounced when $C$ increases. At small shear rates $\gp\lesssim 1$~s$^{-1}$, the deviations from Herschel-Bulkley behaviour would traditionally be attributed to paramount wall slip effects  \cite{Meeker:2004a,Divoux:2011a}. Wall slip is indeed present in our experiments, even though a roughened cell is used, as will be confirmed below in Sect.~\ref{s.creep_rough}. However, the shape of the flow curves of Fig.~\ref{f.rheolC} at low shear rates may also be interpreted as the consequence of time-dependent effects. Indeed CB gels have recently been described as rheopectic systems with a yield stress that depends on previous shear history  \cite{Ovarlez:2013}. For instance the yield stress of a 6\%~w/w CB gel could be made arbitrarily small by applying a succession of steps of decreasing shear stresses. Another signature of such time-dependence is the presence of hysteresis loops in up-down flow curves  \cite{Divoux:2013}.

Therefore, one has to keep in mind that all the rheological features in the present section are strongly time- and protocol-dependent. Note that this remark may be quite general not only in the case of attractive gels but also in the case of glasses, which is usually poorly emphasized in the literature. Here, in particular, the estimates of $\sigma_{y1}$ and $\sigma_{y2}$ depend on the sweep rate used in both oscillatory tests and flow curve determination. Thus, it is not clear whether the shoulders seen in Fig.~\ref{f.rheol}(b) for both $G'$ and $G''$ above $\sigma_{y1}$ at $C=4$ and 6\%~w/w correspond to the two-step yielding of Ref.~ \cite{Koumakis:2011} or not. As shown in Refs.~ \cite{Osuji:2008b,Ovarlez:2013}, both viscoelastic moduli and the yield stress also depend on the stress level during preshear. In other words, the preshear stress determines the gel microstructure and its subsequent mechanical response. Consequently the values reported in Table~\ref{t.rheol} are only relative to the specific protocol used in the present work. This sensitivity on protocol and on previous history also justifies the time-resolved local measurements detailed below that aim at better understanding the slow dynamics involved in yielding.

Finally, we checked that hollow glass microspheres do not have any significant impact on the rheology of our carbon black gels, except for a slight stiffening, as shown in Supplemental Fig.~2 for $C=6$\%~w/w and as already observed in carbopol microgels  \cite{Divoux:2011b}. 

\begin{figure*}[!t]\tt
\centering
\includegraphics[width=0.8\linewidth]{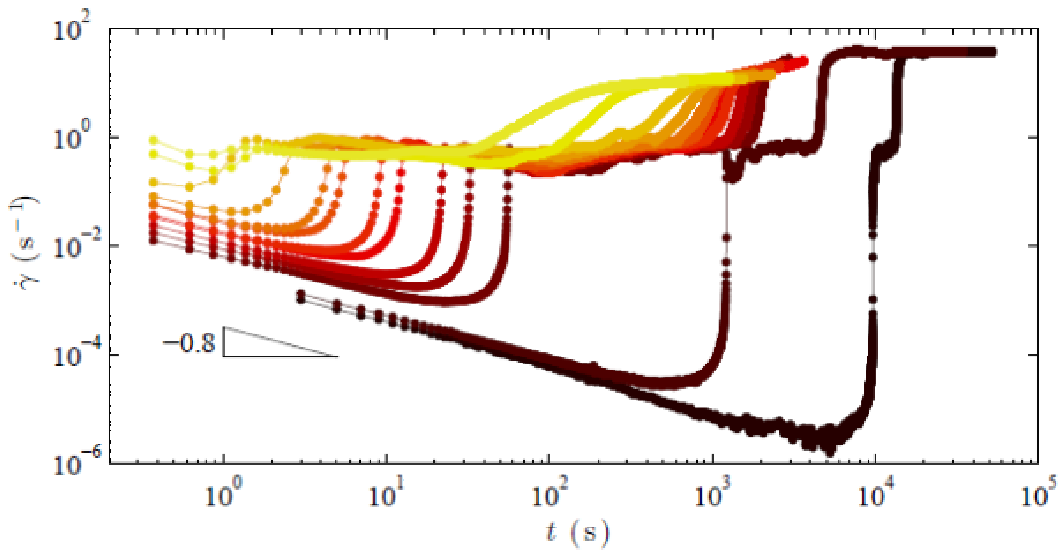}
\caption{Creep experiments in an 8\%~w/w CB gel under rough boundary conditions. Shear rate responses $\gp(t)$ for different shear stresses $\sigma$ applied at time $t=0$: from right to left, $\sigma=24$, 27, 35, 38, 41, 45, 47, 50, 52, 55, 60, 70, and 80~Pa.}
\label{f.gpt_rough}
\end{figure*}

\subsection{Investigation of creep and yielding}
\label{s.creep_protocol}

In order to investigate creep and yielding in our CB samples, we turn to the specific protocol already used in Ref.~ \cite{Gibaud:2010} that allows for reproducible measurements. Prior to each experiment, the CB gel is presheared at +10$^3$~s$^{-1}$ for 20~s and at -10$^3$~s$^{-1}$ for 20~s in order to erase any previous shear history. {In particular, such a high shear rate allows us to exclude structuration effects such as those reported in Ref.~ \cite{Grenard:2011} in smaller gaps. Then small-amplitude} oscillatory shear measurements are performed for 300~s at a frequency of 1~Hz and a stress amplitude of 2~Pa, low enough to be into the linear regime for all samples [see Fig.~\ref{f.rheol}(b)] and long enough for the gel to form {a homogeneous space-spanning network that has reached} a steady state  \cite{Gibaud:2010}. Finally, a constant stress $\sigma$ is applied in the ``positive'' direction, {i.e. in the direction opposite to the last preshear step}, and the resulting shear rate response $\gp(t)$ is recorded by the rheometer simultaneously to the velocity.

Note that this response depends on the preshear protocol, and in particular on the preshearing direction, as shown in Supplemental Fig.~3. Indeed, as recalled above in Sect.~\ref{s.charac}, carbon black gels were shown to be sensitive to a preshear stress $\sigma_p$ through a power-law dependence of the elastic modulus $G'\sim\sigma_p^{1.5-2}$ and through the presence of internal stresses that slowly relax over time  \cite{Osuji:2008b}. Such a complex behaviour may also be interpreted in terms of rheopexy in light of recent results by Ovarlez {\it et al.}  \cite{Ovarlez:2013}. A full investigation of the effect of preshear on yielding is out of the scope of the present article. Still time-dependence and memory effects definitely explain why no consistency can be easily found between the fast sweeps used in the previous section and the long-lasting creep experiments detailed below.

\begin{figure}[!b]\tt
\centering
\includegraphics[width=0.9\columnwidth]{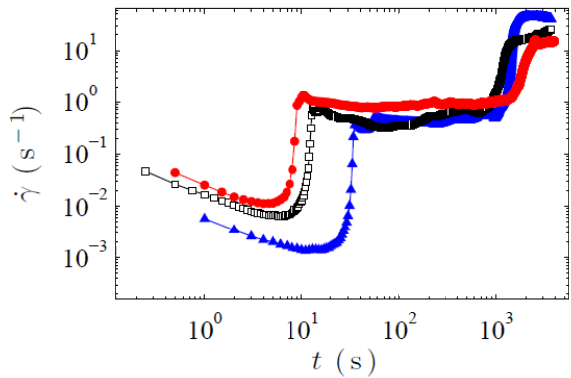}
\caption{Shear rate responses $\gp(t)$ for different CB gels under rough boundary conditions: $C=6$\%~w/w and $\sigma=13$~Pa ($\blacktriangle$), $C=8$\%~w/w and $\sigma=45$~Pa ($\square$), and $C=10$\%~w/w and $\sigma=90$~Pa ($\bullet$). The imposed shear stresses were chosen so as to yield roughly the same failure times.}
\label{f.creepC}
\end{figure}

\section{Experimental results}

\subsection{Yielding scenario under rough boundary conditions}
\label{s.creep_rough}

Figure~\ref{f.gpt_rough} shows the shear rate responses $\gp(t)$ for different shear stresses applied at time $t=0$ on an 8\%~w/w CB gel. Whatever the applied stress $\sigma$, three regimes are observed. (i)~The shear rate first decreases as a power law $\gp(t)\sim t^{-0.8}$ for a duration that strongly decreases as $\sigma$ is increased. This power-law decrease is reminiscent of Andrade creep, also referred to as {\it primary} creep, in solids~ \cite{Andrade:1910}. (ii)~Then $\gp(t)$ progressively departs from power-law creep ({\it secondary} creep regime) and suddenly jumps by several orders of magnitude ({\it tertiary} creep regime) to reach a quasi-constant value $\gp^\star\simeq 0.3$--1~s$^{-1}$. (iii)~Finally $\gp(t)$ undergoes another sharp increase before it reaches a steady-state value. As demonstrated in Fig.~\ref{f.creepC}, the same two-step fluidization process is observed under rough boundary conditions whatever the gel concentration $C$.

\begin{figure*}[!t]\tt
\centering
\includegraphics[width=0.85\linewidth]{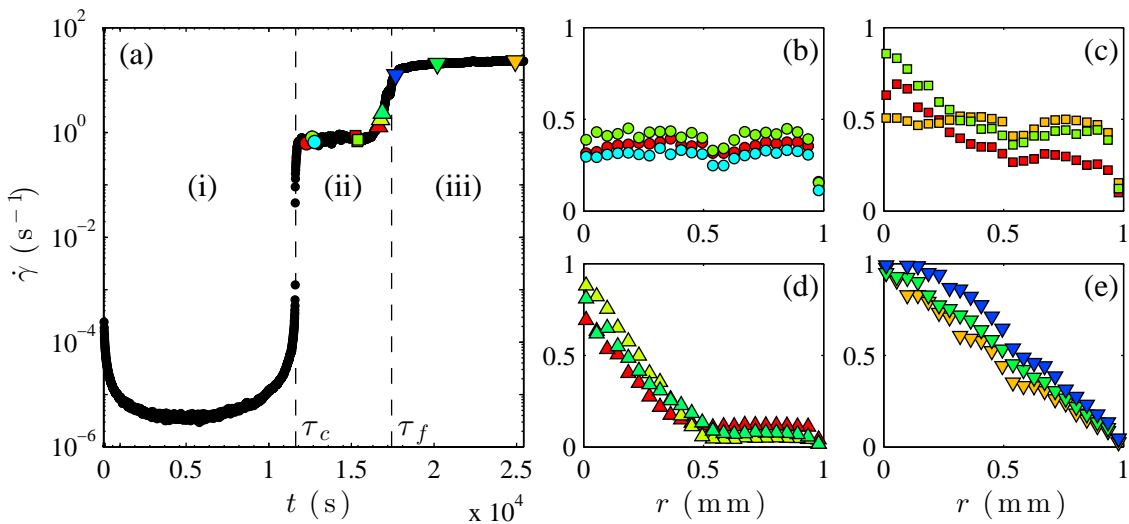}
\caption{Creep experiment in a 10\%~w/w CB gel at $\sigma=55$~Pa under rough boundary conditions. (a) Shear rate response $\gp(t)$. The vertical dashed lines indicate the limits of the three regimes discussed in the text, i.e. the end of the initial creep at $\tau_c=11630$~s and the end of the fluidization regime at $\tau_f=17440$~s. The coloured symbols show the times at which the velocity profiles in (b)--(e) are recorded. Shown in the right panel are individual velocity profiles $v(r,t_0)$, where $r$ is the distance to the rotor, normalized by the current rotor velocity $v_0(t_0)$ at (b)~$t_0=12278$, 12630, and 12780~s, (c)~$t_0=15258$, 15358, and 15401~s, (d)~$t_0=16642$, 16773, and 16882~s, and (e)~$t_0=17696$, 20204, and 24919~s.}
\label{f.velocity_rough}
\end{figure*}

Furthermore, simultaneous velocity measurements allow us to better understand such a temporal evolution. This is illustrated in the case of a 10\%~w/w CB gel in Fig.~\ref{f.velocity_rough}, where $\gp(t)$ is plotted in semi-logarithmic scales. The following sequence of velocity profiles is found. 
During the initial creep regime, noted (i) in Fig.~\ref{f.velocity_rough}(a), velocities are too small to be correctly measured with ultrasonic velocimetry. Therefore, we cannot yet conclude on the strain field during the Andrade-like power-law creep before the first jump in $\gp(t)$ from this experiment. More insights on the creep regime will be given below in the discussion of Sect.~\ref{s.creep}. After the transition to the plateau at $\gp^\star\simeq1$~s$^{-1}$, however, velocity profiles clearly reveal the occurrence of total wall slip at both boundaries in spite of the presence of roughness [Fig.~\ref{f.velocity_rough}(b)]. 

Toward the end of the plateau in $\gp(t)$ [i.e. end of regime (ii) in Fig.~\ref{f.velocity_rough}(a)], pluglike flow gives way to highly fluctuating velocity profiles that oscillate between pluglike and heterogeneous profiles characterized by the coexistence of a highly sheared region and an unsheared band [Fig.~\ref{f.velocity_rough}(c)]. Such fluctuations are most likely due to heterogeneity in the azimuthal direction: the gel is partially broken down into fluidlike and solidlike pieces, which leads to shear-banded profiles that alternate with pluglike profiles. Just before the second inflection point in $\gp(t)$, stable shear-banded velocity profiles are recorded over a rather short time window [Fig.~\ref{f.velocity_rough}(d)]. Finally, at the second inflection point in $\gp(t)$, shear banding evolves into homogeneous flow with no significant wall slip [Fig.~\ref{f.velocity_rough}(e)], which constitutes the steady-state regime noted (iii) in Fig.~\ref{f.velocity_rough}(a). The same sequence of velocity profiles is found in regimes (ii) and (iii) under rough boundary conditions for all gel concentrations $C$.

\subsection{Characteristic timescales under rough boundary conditions}

\subsubsection{Creep duration and fluidization time.~}

\begin{figure}[!t]\tt
\centering
\includegraphics[width=0.8\columnwidth]{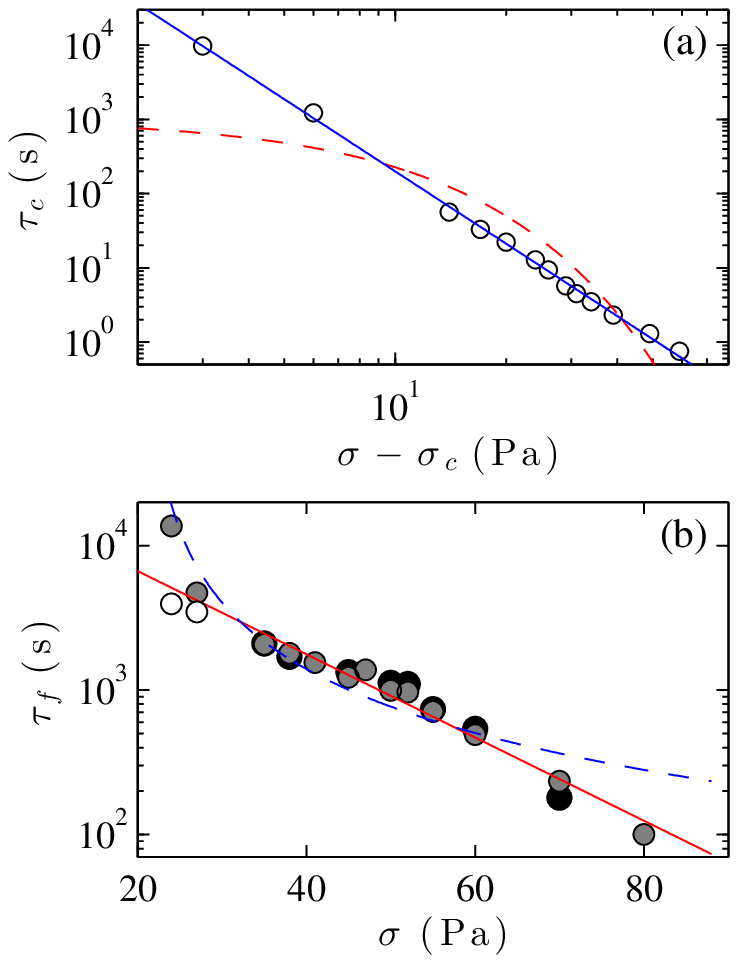}
\caption{Characteristic timescales for yielding of an 8 \%~w/w CB gel under rough boundary conditions as a function of the imposed shear stress $\sigma$. (a) Duration of the creep regime $\tau_c$ (or equivalently time at which total wall slip is observed) as a function of the reduced shear stress $\sigma-\sigma_c$ with $\sigma_c=21.0$~Pa in logarithmic scales. The blue line shows the best power-law fit $\tau_c\sim(\sigma-\sigma_c)^{-\beta}$ with $\beta=3.2\pm 0.1$. The red dashed line is the best exponential fit of $\tau_c$. (b) Full fluidization time $\tau_f$ vs $\sigma$ in semilogarithmic scales. Gray symbols correspond to estimations derived from rheological data while black symbols are times extracted from velocity measurements. White symbols show $\tau_f-\tau_c$ for the two lowest stresses (where $\tau_c>0.2\tau_f$). The red line shows the best exponential fit $\tau_f=\tau_0\exp(-\sigma/\sigma_0)$ with $\tau_0=2.6\pm 0.1\,10^4$~s and $\sigma_0=15.0\pm 0.1$~Pa when excluding the two leftmost points. The blue dashed line is the best power-law fit $\tau_f\sim(\sigma-\sigma_c)^{-\beta}$ of the full data set with $\sigma_c=21.0$~Pa. No satisfactory power-law fit of $\tau_f$ can be found when allowing $\sigma_c$ to vary.}
\label{f.times_rough}
\end{figure}

\begin{figure}[!t]\tt
\centering
\includegraphics[width=0.8\columnwidth]{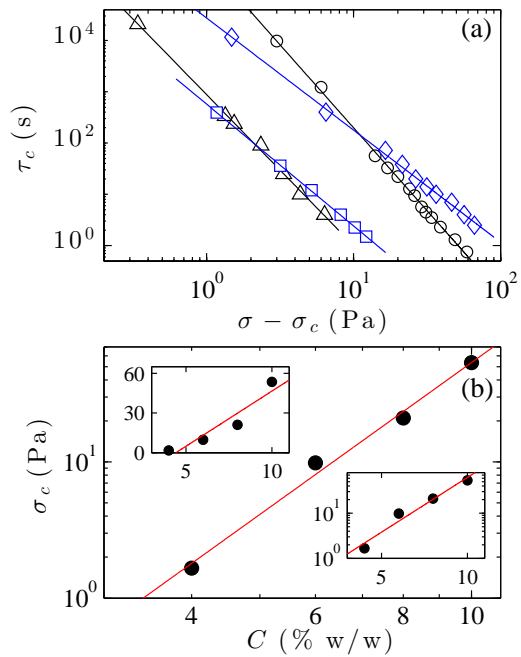}
\caption{(a) Duration of the creep regime $\tau_c$ vs $\sigma$ for different CB gels under rough boundary conditions: $C=4$ ($\triangle$), 6 ($\square$), 8 ($\circ$), and 10\%~w/w~($\diamond$) from left to right. Solid lines show the best power-law fits $\tau_c\sim(\sigma-\sigma_c)^{-\beta}$. The exponents $\beta$ are given in Table~\ref{t.fits}. (b)~Fit parameter $\sigma_c$ as a function of the gel concentration $C$ in logarithmic scales. The red line is the power law $\sigma_c\sim C^{3.7}$. Top (bottom resp.) inset: same data as in the main figure but plotted in linear (semilogarithmic resp.) scales together with the best linear (exponential resp.) fit in red solid line.}
\label{f.tcC_rough}
\end{figure}

\begin{figure}[!t]\tt
\centering
\includegraphics[width=0.8\columnwidth]{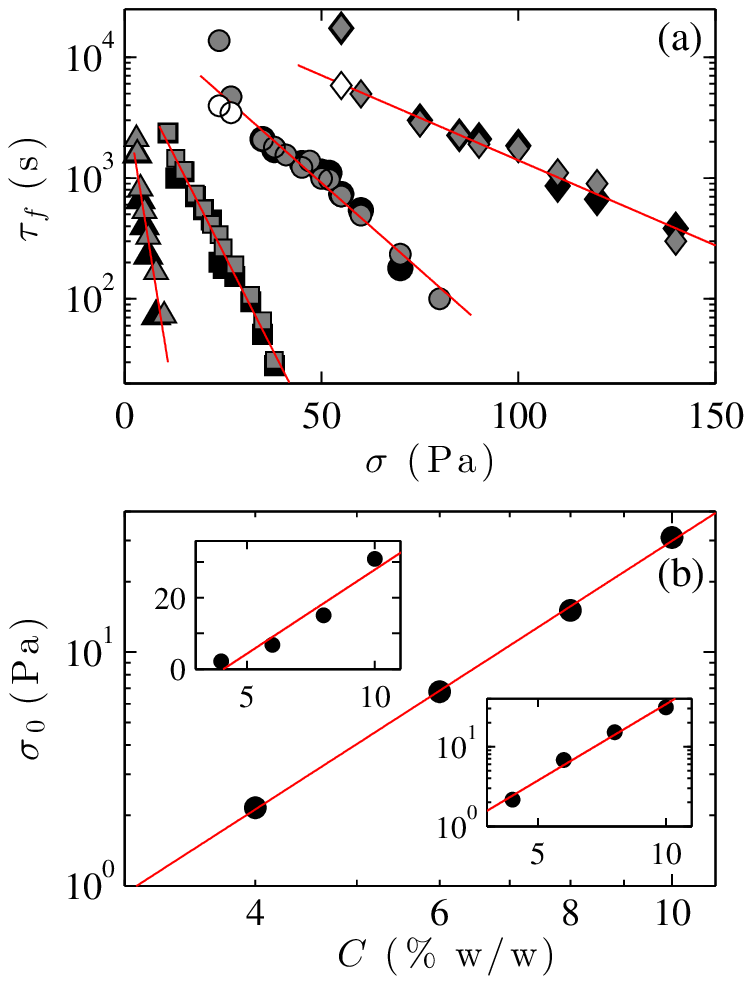}
\caption{(a) Fluidization time $\tau_f$ vs $\sigma$ for different CB gels under rough boundary conditions: $C=4$ ($\triangle$), 6 ($\square$), 8 ($\circ$), and 10\%~w/w~($\diamond$) from left to right. Gray symbols correspond to estimations derived from rheological data while black symbols are times extracted from velocity measurements. The red lines are the best exponential fits $\tau_f=\tau_0\exp(-\sigma/\sigma_0)$ when excluding stresses where $\tau_c>0.2\tau_f$ (the three corresponding points are indicated by showing $\tau_f-\tau_c$ in white symbols). (b)~Fit parameter $\sigma_0$ as a function of the gel concentration $C$ in logarithmic scales. The red line is the power law $\sigma_0\sim C^{2.9}$. Top (bottom resp.) inset: same data as in the main figure but plotted in linear (semilogarithmic resp.) scales together with the best linear (exponential resp.) fit in red solid line.}
\label{f.tfC_rough}
\end{figure}

Based on the above results, we may extract two characteristic times for creep and yielding of CB gels: the time $\tau_c$ that corresponds to the end of the creep regime and to the occurrence of total wall slip and the time $\tau_f$ which corresponds to full fluidization. $\tau_f$ can be estimated either from velocity measurements as the time after which velocity profiles remain all linear, or from the last inflection point in $\gp(t)$. More precisely, we define $\tau_f$ as the inflection time in a linear representation of the shear rate data, i.e. ${\rm d}^2\gp/{\rm d}t^2(\tau_f)=0$. We checked that an estimation of $\tau_f$ from a semilogarithmic representation, i.e. ${\rm d}^2{\rm log}\gp/{\rm d}t^2(\tau_f)=0$, leads to a systematic yet insignificant difference that does not affect our results. Due to the lack of velocity data in the creep regime, $\tau_c$ is only defined from rheological data as the time at which $\gp(t)$ first reaches the plateau at $\gp^\star$. 

Figure~\ref{f.times_rough} shows the two times $\tau_c$ and $\tau_f$ extracted from the data of Fig.~\ref{f.gpt_rough} and from the corresponding velocity profiles. As expected, both times decrease strongly as the imposed shear stress is increased. Interestingly, $\tau_c$ is best fitted by using a power-law $\tau_c\sim(\sigma-\sigma_c)^{-\beta}$ [see Fig.~\ref{f.times_rough}(a)]. The characteristic shear stress $\sigma_c=21.0\pm 0.1$~Pa is first estimated by minimizing the distance to a power-law in a least-square procedure explained in more details in Ref.~ \cite{Divoux:2011a}. The power-law exponent is then found to be $\beta=3.2$. On the other hand, we could not fit $\tau_f$ satisfactorily with a similar power law even by using different estimates for $\sigma_c$. Figure~\ref{f.times_rough}(b) rather shows that the time for full fluidization is well modeled by an exponential behaviour, $\tau_f=\tau_0\exp(-\sigma/\sigma_0)$, if one excludes the data point at the lowest shear stress. This result is fully consistent with previous works  \cite{Gibaud:2010,Sprakel:2011,Lindstrom:2012}. Also note that, except for the longest failure times [see, e.g., Fig.~\ref{f.velocity_rough}(a)], $\tau_f$ remains always much larger than $\tau_c$ so that the duration of the fluidization process {\it after} the initial creep, $\tau_f-\tau_c\simeq \tau_f$, also decreases exponentially with the imposed stress. Actually, if $\tau_f-\tau_c$ is considered rather than $\tau_f$ when the creep duration $\tau_c$ becomes a significant fraction of the full fluidization time $\tau_f$ (say when $\tau_c>0.2\tau_f$) at low shear stress, one recovers an exponential behaviour all the way down to the lowest shear stresses as shown by the two white symbols in Fig.~\ref{f.times_rough}(b) [see also Fig.~\ref{f.tfC_rough}(a) for the case of the 10\%~w/w CB gel at $\sigma=55$~Pa shown in Fig.~\ref{f.velocity_rough}]. These two different scalings for $\tau_c$ and $\tau_f-\tau_c$ (or equivalently $\tau_f$ in most cases) will be further discussed below in Sect.~\ref{s.creepdur} and \ref{s.model} in light of other experiments and recent modeling of yielding.

\subsubsection{Influence of the gel concentration.~}
\label{s.timesC}

The influence of the gel weight fraction $C$ on the creep duration $\tau_c$ and on the fluidization time $\tau_f$ is summarized in Figs.~\ref{f.tcC_rough} and \ref{f.tfC_rough} respectively. Figure~\ref{f.tcC_rough}(a) confirms that whatever the concentration $\tau_c$ behaves as a power law of a ``viscous'' stress $\sigma-\sigma_c$, where the critical stress $\sigma_c$ depends on the concentration $C$ as a power law $\sigma_c\sim C^{3.7}$ [see Table~\ref{t.fits} and Fig.~\ref{f.tcC_rough}(b)]. The insets of Fig.~\ref{f.tcC_rough}(b) show that in spite of the limited range of concentration that only spans half a decade, both affine and, {to a lesser extent}, exponential dependences for $\sigma_c$ vs $C$ {are unlikely} since a systematic curvature of the data is seen in linear and semilogarithmic coordinates. The power law $\sigma_c\sim C^{3.7}$ is fully consistent with that of the rheological parameters $G'$, $G''$,  and $\sigma_{y2}$ (see Table~\ref{t.rheol} and Fig.~\ref{f.rheolC}).

\begin{table}[h!]
\small
  \caption{\small Fit parameters vs gel concentration $C$ for $\tau_c$ and $\tau_f$, the two timescales involved in yielding of CB gels under rough boundary conditions: $\tau_c\sim(\sigma-\sigma_c)^{-\beta}$ and  $\tau_f=\tau_0\exp(-\sigma/\sigma_0)$. The last line shows the exponents extracted from the power-law fits shown in Figs.~\ref{f.tcC_rough}(b) and \ref{f.tfC_rough}(b).}
  \label{t.fits}
  \begin{tabular*}{0.49\textwidth}{@{\extracolsep{\fill}}llllll}
    \hline
     $C$ & $\sigma_c$ (Pa) &  $\beta$ & $\tau_{0}$ ($\times 10^4$~s) & $\sigma_0$ (Pa)  \\
    \hline
    4\% &  1.7         & 2.9 & 0.61 & 2.0\\
    6\% &  9.8	& 2.4  & 1.1 & 6.6\\
    8\% &   21.0 	& 3.2 & 2.6 & 15.0\\
    10\% &  53.5	& 2.2 & 3.8 &30.4\\
    \hline
exp. & 3.7 & n/a & n/a & 2.9\\
  \hline
  \end{tabular*}
\end{table}

Turning to the fluidization time $\tau_f$, as already visible in Fig.~\ref{f.times_rough}(b), the estimations of $\tau_f$ from rheological data (gray symbols) and from velocity measurements at a given height in the Couette cell (black symbols) are in very good agreement [see Fig.~\ref{f.tfC_rough}(a)]. This allows us to reconcile the two independent studies by Gibaud {\it et al.} \cite{Gibaud:2010} and Sprakel {\it et al.} \cite{Sprakel:2011} which were based on these two different definitions of $\tau_f$. This also suggests that the fluidization process occurs rather homogeneously along the vertical direction, although this remains to be directly checked using two-dimensional imaging techniques  \cite{Gallot:2013}. Moreover, as expected from first intuition, $\tau_f$ strongly increases with $C$ for a given $\sigma$. Equivalently the stress required to fully fluidize a gel after a given time $\tau_f$ dramatically increases with $C$. Whatever the concentration an exponential law fits the stress-dependence of $\tau_f$ (or of $\tau_f-\tau_c$) very well. The characteristic stress $\sigma_0$ involved in this exponential decay is plotted against $C$ in Fig.~\ref{f.tfC_rough}(b). The evolution of $\sigma_0$ with $C$ is best modeled by a power law $\sigma_0\sim C^{2.9}$ in spite of a small range of concentrations [see also Table~\ref{t.fits} and insets of Fig.~\ref{f.tfC_rough}(b)]. This exponent of 2.9 is close to that found for the concentration-dependence of the yield stresses $\sigma_{y1}$ but significantly smaller than that of $G'$, $G''$,  and $\sigma_{y2}$ vs $C$ (see Table~\ref{t.rheol} and Fig.~\ref{f.rheolC}). A discussion on this power-law dependence and on the value of the exponent will be provided in Sect.~\ref{s.model}.

\subsection{Comparison with smooth boundary conditions}
\label{s.bc}

\begin{figure*}[!t]\tt
\centering
\includegraphics[width=0.8\linewidth]{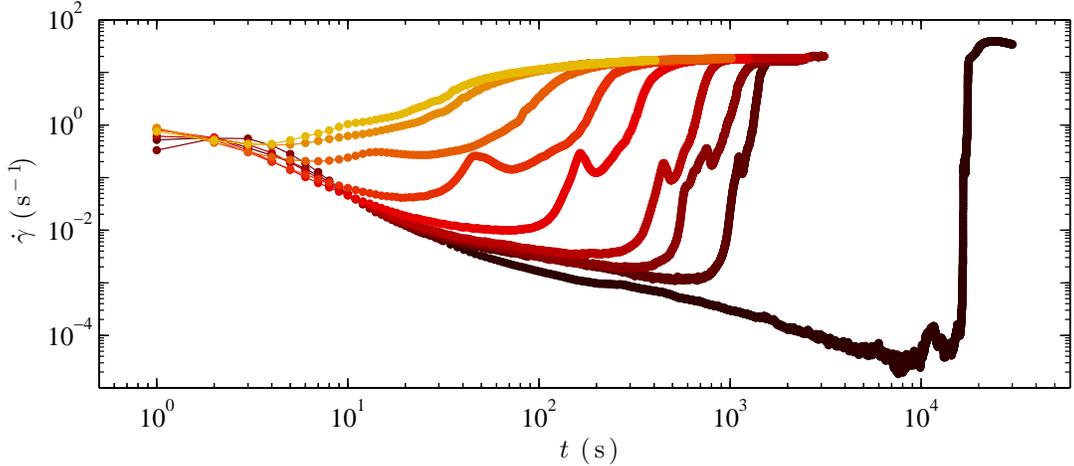}
\caption{Creep experiments in an 8\%~w/w CB gel under smooth boundary conditions. Shear rate responses $\gp(t)$ for different shear stresses $\sigma$ applied at time $t=0$: from right to left, $\sigma=45$, 50, 52, 53, 55, 60, 65, 70, and 75~Pa.}
\label{f.gpt_smooth}
\end{figure*}

Yielding under smooth boundary conditions has been investigated by Gibaud {\it et al.}  \cite{Gibaud:2010} in the case of a 6\%~w/w CB gel. Figure~\ref{f.gpt_smooth} shows a set of creep responses recorded in an 8\%~w/w CB gel. {Overall} the shear rate response under smooth boundary conditions {resembles} that under rough boundary conditions. The shear rate first decreases in a long creeping flow regime before undergoing an upturn leading to a fluidized steady state. {However here the sudden jump in $\gp(t)$ does not lead to such a well-defined plateau as under rough boundaries but rather to one (or several) kink(s) and fluctuations in the shear rate before a final increase up to steady state (see also Supplemental Fig.~4 for a semilogarithmic representation of $\gp(t)$).} Therefore three regimes can no longer be clearly distinguished in $\gp(t)$ and one may only extract a single characteristic time $\tau_f$ for yielding and fluidization, defined as the last inflection point of $\gp(t)$. Moreover the creep regime is not characterized by a well-defined power-law decay of $\gp(t)$ as in the case of rough boundaries.

As already reported in Ref.~ \cite{Gibaud:2010} velocity profiles show an evolution that is quite similar to that reported above for rough boundary conditions. Still, with smooth walls, apparent shear rates are larger, which makes velocity measurements during the initial creep regime possible. Such measurements show that the sample undergoes total slippage right from the start-up of shear at $t=0$ (see Supplemental Fig.~4 and Fig.~3 in Ref.~ \cite{Gibaud:2010}). Total wall slip persists until the fluctuations seen in $\gp(t)$ for intermediate shear stresses (see, e.g., $\sigma=50$--55~Pa in Fig.~\ref{f.gpt_smooth}). These fluctuations signal the beginning of bulk fluidization through highly fluctuating shear-banded velocity profiles as shown in Supplemental Fig.~4. In steady state, a small amount of wall slip remains measurable at both walls contrary to the case of rough boundaries.

\begin{figure}[!t]\tt
\centering
\includegraphics[width=0.8\columnwidth]{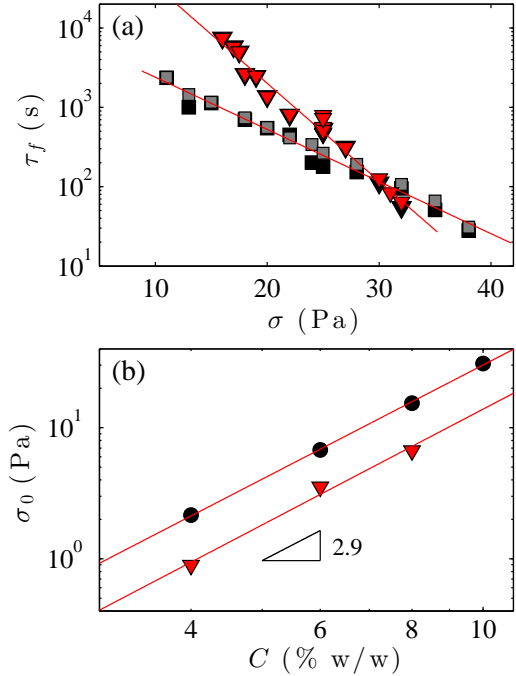}
\caption{(a) Fluidization time $\tau_f$ vs $\sigma$ for a 6\%~w/w CB gel under rough ($\square$) and smooth ($\triangledown$) boundary conditions. Gray and red symbols correspond to estimations derived from rheological data while black symbols are times extracted from velocity measurements. Red lines are the best exponential fits $\tau_f=\tau_0\exp(-\sigma/\sigma_0)$. (b)~Fit parameter $\sigma_0$ as a function of the gel concentration $C$ in rough ($\bullet$) and smooth ($\triangledown$) boundary conditions. Red lines are power laws $\sigma_0\sim C^{2.9}$.}
\label{f.tfBC}
\end{figure}

Fluidization times in smooth and rough Couette cells are compared in Fig.~\ref{f.tfBC}(a) in the case of a 6\%~w/w CB gel. In both cases, $\tau_f$ follows an exponential behaviour with the imposed shear stress. However, at low imposed stresses, $\tau_f$ is seen to be larger for smooth boundaries while it becomes smaller than in the case of rough walls at high stresses. In other words, the exponential decay is stronger in the case of smooth boundary conditions. This is also observed for other concentrations: the parameter $\sigma_0$ remains always smaller for smooth boundary conditions as shown in Fig.~\ref{f.tfBC}(b). Since the 10\%~w/w CB gel is seen to slip on smooth walls whatever the imposed stress, only three concentrations are reported in Fig.~\ref{f.tfBC}(b). Still the data are again compatible with a power-law behaviour $\sigma_0\sim C^{2.9}$ with a prefactor that is about twice as small in the case of smooth walls.

\section{Discussion}
\label{s.discuss}

\subsection{Andrade-like creep regime}
\label{s.creep}

In this paragraph, we discuss the strikingly robust initial creep regime found with rough boundary conditions and characterized by a power-law decay of the shear rate as $\gp\sim t^{-0.8\pm 0.1}$ or equivalently by an increase of the strain as $\gamma\sim t^{0.2\pm 0.1}$. Such a power-law creep is reminiscent of the Andrade law $\gp\sim t^{-2/3}$ (or $\gamma\sim t^{1/3}$) found in solid materials  \cite{Andrade:1910} and has been attributed to
collective dislocation dynamics  \cite{Miguel:2002,Csikor:2007,Miguel:2008}. Experiments on heterogeneous fiber materials as well as corresponding models such as the ``fiber bundle model'' (FBM) show a similar creep behaviour prior to rupture  \cite{Kun:2003,Nechad:2005a,Rosti:2010,Jagla:2011}. Andrade-like creep has also been reported for cellulose gels  \cite{Plazek:1960} and more recently for some amorphous soft solids such as polycrystalline surfactant hexagonal phases  \cite{Bauer:2006}, carbopol microgels  \cite{Divoux:2011a}, core-shell poly(styrene)-poly(N-isopropylacrylamide) colloidal particles  \cite{Siebenburger:2012}, and thermo-reversible protein gels  \cite{Brenner:2013}. Yet, no clear link between the physical mechanisms at play in the creep of ordered and disordered materials is available.

\begin{figure}[!t]\tt
\centering
\includegraphics[width=0.8\columnwidth]{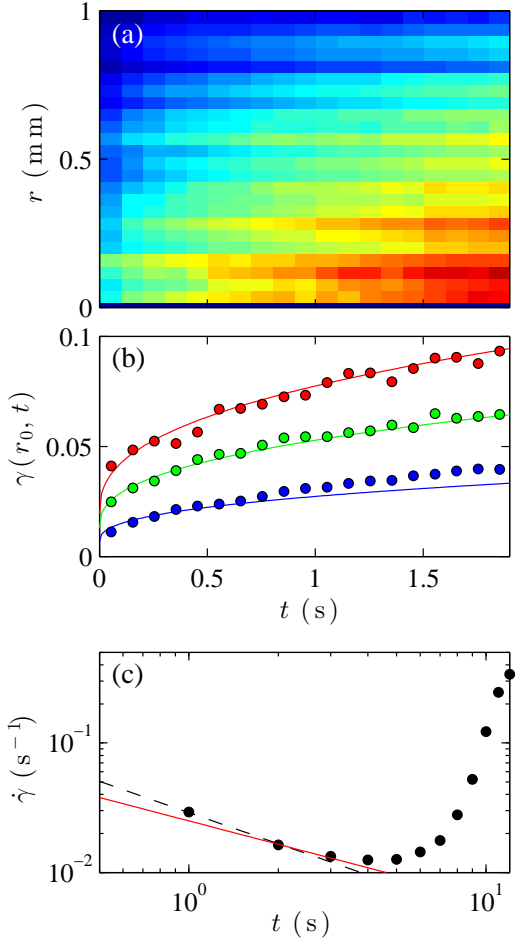}
\caption{Creep regime for a 6\%~w/w CB gel at $\sigma=17$~Pa under rough boundary conditions. (a) Spatiotemporal diagram of the local strain $\gamma(r,t)$ coded in linear color levels. Blue corresponds to 0 and red to a strain of 0.1. (b) Local strain $\gamma$ versus time $t$ at various locations within the gap: $r_0=0.12$ (top red symbols), 0.40 (middle green symbols), and 0.69~mm (bottom blue symbols). The solid lines are $\gamma(r_0,t)=(\gamma_0+\lambda t^{0.4})\cdot (1- r_0/e)$ with $\gamma_0=2.2$~\% and $\lambda=0.066$. (c) Global shear rate response $\gp(t)$ recorded simultaneously by the rheometer. The black dotted line and the red solid line are power laws with exponents -0.8 and -0.6 respectively.}
\label{f.creep}
\end{figure}

In order to get a deeper insight into the creep regime, Fig.~\ref{f.creep} focuses on an experiment where strains in the creep regime are large enough to be estimated from ultrasonic velocimetry. This results from a difficult compromise between measurable motions (large applied stress) and long enough creep regime (small applied stress). The local strain $\gamma(r,t)$ is computed {from the ultrasonic data and independently of any rheological measurement} by summing up the local displacements computed from the cross-correlation algorithm described in Ref.~ \cite{Manneville:2004a}. The space-time map of $\gamma(r,t)$ shown in Fig.~\ref{f.creep}(a) suggests that the strain increases slowly but mostly homogeneously after an initial instantaneous elastic deformation $\gamma_0$. Indeed in view of the large uncertainty on $\gamma$, it seems reasonable to attribute the observed fluctuations of $\gamma$ with $r$ to increased noise due to smaller local intensity of the ultrasonic signal and/or tiny displacements rather than to any true spatial heterogeneity of the strain field. The scenario of a bulk deformation that remains homogeneous at least on the length scales probed by our high-frequency ultrasonic setup (typically 1~$\mu$m) is further supported by Fig.~\ref{f.creep}(b) which shows that the local strain is well captured by $\gamma(r,t)=(\gamma_0+\lambda t^{0.4})\cdot (1- r/e)$ with the same set of parameters $\gamma_0$ and $\lambda$ for three different positions within the gap. The exponent 0.4 found here is slightly larger than expected from global rheology but, in this specific case, the creep regime is short ($\tau_c\simeq 10$~s) and the exponent of the global shear rate $\gp(t)$ vs $t$ actually decreases continuously from -0.8 at very early times to 0 when the minimum is reached, so that an exponent of 0.4 is not inconsistent with some average exponent derived from global rheology [see red line with slope -0.6 in Fig.~\ref{f.creep}(c)]. This point clearly deserves more attention and should be addressed in detail in future work together with the fact that the scaling behaviour of $\gamma(r,t)$ may also be weakly space-dependent as suggested by the systematic deviations from the power law close to the stator [see data in blue in Fig.~\ref{f.creep}(b)].

\begin{figure}[!t]\tt
\centering
\includegraphics[width=0.8\linewidth]{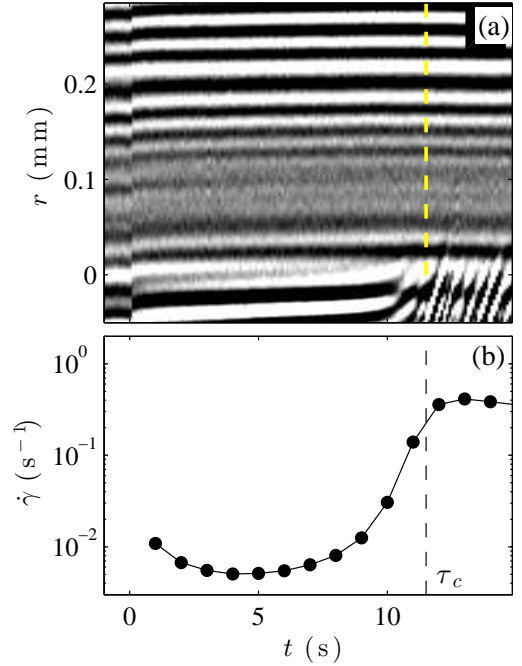}
\caption{Creep regime for a 6\% w/w CB gel at $\sigma=15$~Pa under rough boundary conditions. (a)~Spatiotemporal diagram of the pressure signal plotted as a function of the radial position $r$ from the rotor (vertical axis) and time $t$ (horizontal axis) after shear is started at $t=0$. Negative values of $r$ correspond to locations inside the Plexiglas rotor. (b)~Corresponding shear rate response $\gp(t)$. The dashed lines indicate the end of the creep regime at $\tau_c\simeq 11.5$~s which is seen to correspond to wall slip at the rotor.}
\label{f.creep_speckle}
\end{figure}

{Figure~\ref{f.creep_speckle}(a) investigates the initial creep at a slightly lower shear stress by looking directly at the ultrasonic speckle signal. It focuses on the region close to the rotor and extends slightly within the rotor ($-0.05<r<0.28$~mm). The sharp shift in the pressure signal at $t=0$ corresponds to instantaneous elastic response when shear is started. It is followed by a much slower deformation over roughly the first 10~s. During this Andrade-like creep regime, no obvious heterogeneity is observed in the bulk. For $t\gtrsim 10$~s, the rotor progressively accelerates while pressure signals remain horizontal in the bulk. In other words, the material starts to slip at the rotor. For $t\gtrsim 12$~s, the rotor has achieved a large, steady value as indicated by the constant slope of the ultrasonic echoes inside the rotor (for $-0.05<r<0$~mm). Figure~\ref{f.creep_speckle}(b) confirms that the time $\tau_c$ that signals the end of the creep regime also corresponds to full wall slip at the rotor.}

{Therefore both Figs.~\ref{f.creep} and \ref{f.creep_speckle} show that no bulk fracture nor any noticeable local rearrangements are observed} before the material detaches from the inner wall. This means that, if present, bulk plasticity during the Andrade creep regime would involve motions on length scales that are below the detection threshold of our ultrasonic technique (typically 1~$\mu$m). Another possibility is that plastic events are preferentially localized close to the moving wall. Indeed one may be tempted to interpret the somewhat larger fluctuations of $\gamma(r,t)$ close to the rotor [see data in red in Fig.~\ref{f.creep}(b)] as an indication for such plasticity localization. Yet, we again emphasize that the poor signal-to-noise ratio prevents us from drawing any definite conclusion at this stage. {In spite of the good temporal resolution of the present measurements (0.1~s in the case of the ultrasonic data of Figs.~\ref{f.creep} and \ref{f.creep_speckle}), an apparently homogeneous strain field could also result from an average over many plastic events occurring on timescales much shorter than our measurement time. Note that a similar} homogeneous flow was also found during the Andrade-like creep regime evidenced in carbopol microgels  \cite{Divoux:2011a} although in this case the flow profile was averaged over a long duration in the creep regime. We also recall that in both carbopol and carbon black gels we actually follow the motion of acoustic tracers and not that of the material itself. Therefore more direct local measurements \cite{Seth:2012} using, e.g. fast confocal microscopy, are required to conclude on the physical origin of the creep regime.
 
\subsection{Duration of the creep regime: insights from fiber-bundle models (FBMs)}
\label{s.creepdur}

One of the main results of the present work is the observation of a two-step yielding process under rough boundary conditions involving two timescales, $\tau_c$ and $\tau_f$, that show very different dependences upon the imposed shear stress. We first discuss the duration $\tau_c$ of the creep regime, which is found to scale as $\tau_c\sim(\sigma-\sigma_c)^{-\beta}$. Such a power-law behaviour is also found for the ``rupture time'' in various FBMs that aim at reproducing systems with a succession of Andrade creep (primary creep), quasi stationary regime (secondary creep), and acceleration of the strain rate (tertiary creep) prior to rupture  \cite{Kun:2003,Nechad:2005a,Jagla:2011,Nechad:2005b}. Depending on the fiber rheology and on the relaxation mechanisms after fiber breakage, FBM models predict $\beta=0.5$--1.25  \cite{Jagla:2011,Nechad:2005b} significantly below our experimental observations for $\tau_c$ in CB gels for which $\beta=2.2$--3.2 (see Table~\ref{t.fits}).

\begin{figure}[!t]\tt
\centering
\includegraphics[width=0.8\columnwidth]{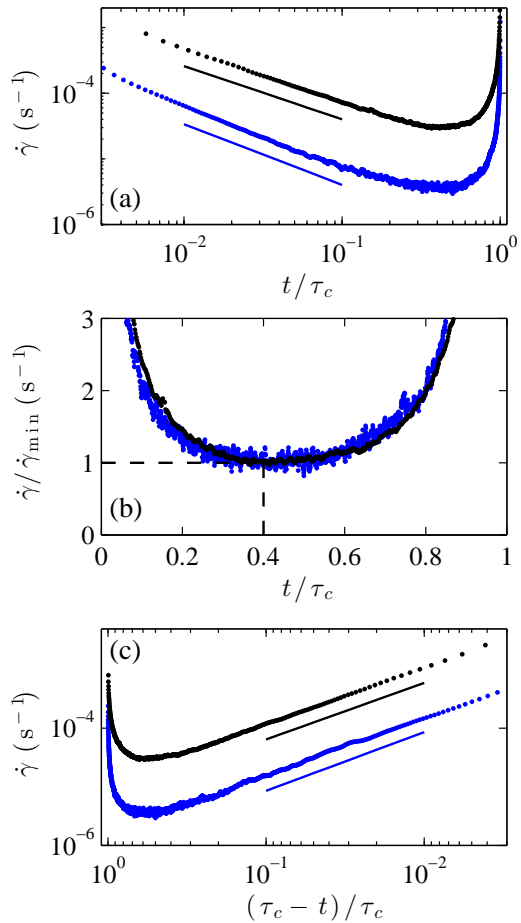}
\caption{Initial creep regimes in two different CB gels already shown in Figs.~\ref{f.gpt_rough} and \ref{f.velocity_rough}. Black symbols correspond to the 8\%~w/w gel at $\sigma=27$~Pa with $\tau_c=1216$~s and blue symbols correspond to the 10\%~w/w gel at $\sigma=55$~Pa with $\tau_c=11575$~s. (a) $\gp$ vs $t/\tau_c$ in logarithmic scales showing the primary (Andrade) creep regime. Solid lines show the best power-law fits (shifted vertically for clarity) for $0.01<t/\tau_c<0.1$ with exponents -0.81 (top) and -0.93 (bottom). (b) $\gp/\gp_{\rm min}$ vs $t/\tau_c$ in linear scales, where $\gp_{\rm min}$ is an estimate of the minimum value reached by $\gp$. The vertical dashed line shows $\tau_{\rm min}=0.4\tau_c$. (c) $\gp$ vs $(\tau_c-t)/\tau_c$ in logarithmic scales (with reverse horizontal axis) showing the tertiary creep regime prior to failure at the inner wall. Solid lines show the best power-law fits (shifted vertically for clarity) for $0.01<(\tau_c-t)/\tau_c<0.1$ with exponents -0.97 (top) and -1.0 (bottom).}
\label{f.regimes}
\end{figure}

\begin{figure}[!t]\tt
\centering
\includegraphics[width=0.75\columnwidth]{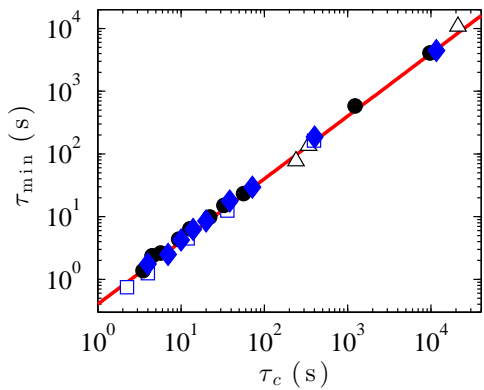}
\caption{Time $\tau_{\rm min}$ at which $\gp(t)$ reaches a minimum during the creep regime plotted against the creep duration $\tau_c$ for CB gels at $C=4$ ($\triangle$), 6 ($\square$), 8 ($\bullet$), and 10\%~w/w~($\blacklozenge$). The red solid line is $\tau_{\rm min}=0.4\tau_c$.}
\label{f.tmin}
\end{figure}

Inspired by these previous works, we investigate in more detail the shear rate responses for $t<\tau_c$ in two different CB gels at 8\%~w/w and 10\%~w/w in Fig.~\ref{f.regimes}. Power-law creep, with exponents $-0.81$ and $-0.93$ in the two particular cases shown in Fig.~\ref{f.regimes}(a), is observed from the earliest stages until $t\simeq 0.2\tau_c$. From $t\gtrsim 0.2\tau_c$, the shear rate progressively deviates from the initial power law and eventually reaches a minimum $\gp_{\rm min}$ at $\tau_{\rm min}\simeq 0.4\tau_c$. The two shear rate responses nicely collapse when normalized by $\gp_{\rm min}$ and plotted against $t/\tau_c$ [Fig.~\ref{f.regimes}(b)]. This clearly indicates that the nature of the creep regime remains the same for these two different concentrations as already inferred from Fig.~\ref{f.creepC}. This is further confirmed in Fig.~\ref{f.tmin} where $\tau_{\rm min}$ is reported as a function of $\tau_c$ for all the creep experiments where $\tau_{\rm min}$ was large enough to be measured (typically larger than 1~s), thus involving the four different gel concentrations. Whatever the CB concentration and the applied stress, the new timescale $\tau_{\rm min}$ is seen to be directly proportional to $\tau_c$ with a prefactor of 0.4. Finally, during the approach to the localized failure at the inner wall, the shear rate increases as a power law of $(\tau_c-t)$ with exponents than can hardly be distinguished from $-1$ (namely $-0.97$ and $-1.0$ respectively). 

Figure~\ref{f.regimes} reports results that are strikingly similar to those of FBM models  \cite{Kun:2003,Jagla:2011} and to experiments on fiber composite materials  \cite{Nechad:2005a,Nechad:2005b}. For instance, the exact same scaling for the power-law acceleration of the strain rate prior to rupture, $\dot\varepsilon\sim1/(\tau_c-t)$, is generally reported. In particular, Fig.~2 in the recent work by Jagla  \cite{Jagla:2011} also shows the three different stages in the creep regime with a minimum reached at a time that is proportional to the final rupture time. Yet the prefactors reported in Refs.~ \cite{Nechad:2005a,Jagla:2011,Nechad:2005b}, typically 0.5--0.7 are significantly larger than that reported here. 

Finally, we note that two types of FBM models have been proposed in the literature: {models that simply rely on a local yield strain (or stress) for fiber rupture} \cite{Nechad:2005a,Jagla:2011,Nechad:2005b}, which lead to rupture times that decrease as a power-law of the viscous stress $\sigma-\sigma_c$, and models {that also include damage accumulation in the form of a memory term involving the whole loading history of the fiber} \cite{Kun:2006,Kun:2008,Halasz:2012}, which lead to a power law of the applied shear stress (i.e. $\sigma_c=0$) otherwise known as the Basquin law. Interestingly, our results are fully consistent with the first category of FBM models. This supports the microscopic picture of attractive gels controlled by a local yield strain between two neighbouring colloidal particles rather than by a local yield stress. This may also explain why kinetic models such as the one developed in Ref. \cite{Sprakel:2011,Lindstrom:2012} do not apply to the duration of the creep regime.

\subsection{Total wall slip and lubrication layers after failure}
\label{s.slip}

As shown in Fig.~\ref{f.velocity_rough}, the succession of primary, secondary, and tertiary creep regimes [noted (i) in Fig.~\ref{f.velocity_rough}(a)] gives way to a regime of total wall slip before fluidization occurs at $\tau_f$ [see Fig.~\ref{f.velocity_rough}(b)]. Here, local velocity measurements prove crucial to interpret the shear rate response under rough boundary conditions, which systematically shows a plateau at a characteristic apparent shear rate $\gp^\star$ right after failure occurs at $\tau_c$. Indeed, along this plateau, the flow profile points to local vanishing shear rates in the bulk, i.e. the gel undergoes solid-body rotation and the flow is pluglike. Therefore, shear is localized in lubrication layers which thickness $\delta_l$ is much smaller than the spatial resolution of our velocity profiles. These lubrication films are submitted to a local shear rate which is of the order of $\gp_l\simeq \gp^\star e/\delta_l$ where $e=1$~mm is the gap of the Couette cell. Assuming that the lubrication layers remain Newtonian with a viscosity $\eta_l$ close to that of the pure suspending mineral oil, one gets $\delta_l\simeq\eta_l\gp^\star e/\sigma$. Quantitatively, the data of Fig.~\ref{f.velocity_rough} leads to $\delta_l\simeq 400$~nm, which is consistent with data reported for surfactant systems \cite{Salmon:2003d} but significantly larger than the thickness of lubrication films found in microgel pastes  \cite{Meeker:2004b,Seth:2008}. {This estimate is however in good agreement with Ref.~ \cite{Kalyon:2005} that reports $\delta_l\simeq (1-\phi/\phi_m)D_p$ for various dispersions of rigid particles, where $\phi$ is the volume fraction, $\phi_m$ the maximum packing fraction, and $D_p$ the particle diameter, since in our case we have $\phi/\phi_m\simeq 0.1$--0.2 and $D_p\simeq 200$--500~nm.}

As reported in Supplemental Fig.~5, $\gp^\star$ increases roughly linearly with the applied stress $\sigma$, which suggests that the size of lubricating layers is independent of $\sigma$ (although $\eta_l$ could also depend on $\sigma$ in the case where a stress-dependent amount of CB particles remains in the lubricating layers after failure at the walls). Supplemental Fig.~5 also indicates that $\gp^\star$ increases with the gel concentration $C$, pointing to thicker films (or to less viscous films) in more concentrated samples. 

Our measurements show that wall slip is inherent to the yielding mechanism of our attractive gels since standard procedures to minimize slippage through wall roughness fail. {However, a more detailed and quantitative study of the influence of roughness would be interesting to better understand wall slip in relation with yielding. In particular, the typical roughness of our sand-blasted cell (1~$\mu$m) lies above the size of the fractal particles (0.2--0.5~$\mu$m) but below that of aggregates in the system at rest. Therefore our ``rough'' boundaries may appear as smooth boundaries for large aggregates but not for smaller aggregates or for individual soot particles. This may partly explain why wall slip is transiently observed under ``rough'' boundaries and account for the concentration dependence observed above since the aggregate size is likely to depend on $C$. Other physico-chemical factors, such as wetting properties or wall--particle electrostatic interactions, most probably influence failure at the wall as well.}

\subsection{Fluidization time: activated dynamics?}
\label{s.model}

The second timescale involved in the yielding of carbon black gels, namely the time $\tau_f$ required for full fluidization, was already shown to follow an exponential decay with the applied shear stress in previous works  \cite{Gibaud:2010,Sprakel:2011,Lindstrom:2012}. Such a behaviour, which was also reported in silica and polystyrene particulate gels \cite{Sprakel:2011,Lindstrom:2012} as well as thermo-reversible protein gels  \cite{Brenner:2013}, hints to activated dynamics with an energy barrier that decreases linearly with the applied stress. Note that a similar exponential decay was found in thermo-reversible silica gels \cite{Gopalakrishnan:2007} and for the shear-induced aggregation time in some non-Brownian suspensions  \cite{Guery:2006}. A mean-field model for delayed yielding was recently proposed in which macroscopic failure results from a homogeneous degradation due to microscopic strand fractures within the gel  \cite{Sprakel:2011,Lindstrom:2012}.

The present experiments shed new light on the validity and application of such a model. Indeed, the model proposed by Lindstrom {\it et al.}~ \cite{Lindstrom:2012} does obviously not apply to the creep duration $\tau_c$ which is found to follow a power-law decay rather than an exponential decay. This is in apparent contradiction with the statement in Ref.~ \cite{Lindstrom:2012} that ``the delay time is well estimated by the time-scale of the initial static fatigue, occurring prior to the final macroscopic failure, while the duration of critical crack propagation, which is much more rapid, can be neglected,'' which would imply that the model applies to the initial creep regime and thus to $\tau_c$. However, one should keep in mind that the experiments reported in Refs.~ \cite{Gibaud:2010,Sprakel:2011,Lindstrom:2012} were restricted to {\it smooth} boundary conditions where slippage is likely to occur at the earliest stages as shown in Sect.~\ref{s.bc} above (see also Supplemental Fig.~4). These experiments show only one timescale $\tau_f$ that decreases exponentially with the applied stress. This suggests to apply models based on activated dynamics such as that devised by Lindstrom {\it et al.}~ \cite{Lindstrom:2012} only for the part of the dynamics where the material has failed at the inner cylinder (i.e. for $t>\tau_c$ in the presence of rough walls and for $t>0^+$ when smooth walls are used). Indeed, as noted above in Sect.~\ref{s.timesC}, under rough boundary conditions, the duration of the fluidization process $\tau_f-\tau_c$ follows a nice exponential decay over the full range of accessible shear stresses [see Fig.\ref{f.tfC_rough}(a)]. We recall that considering $\tau_f-\tau_c$ rather than simply $\tau_f$ compensates for the deviations of $\tau_f$ from exponential as the shear stress approaches the critical stress $\sigma_c$ and $\tau_c$ is no longer negligible when compared to $\tau_f$. This clearly indicates a crossover to a regime that is dominated by the (diverging) timescale for creep $\tau_c$. 

Focusing on the fluidization process, we note that heterogeneous flows are observed over a rather narrow time window close to $\tau_f$. Therefore, under both smooth and rough boundary conditions, the gel remains solid and subject to friction from the lubrication layers at the walls over most of the fluidization regime. If one applies the model of Refs.~ \cite{Sprakel:2011,Lindstrom:2012} to our case where no previous flow history is applied to the sample, then one expects the characteristic stress to be given by $\sigma_0=\rho_0 k_B T/\delta$ where $\rho_0\sim 1/\xi^2$ is the initial area density of strands, with $\xi$ the mesh size (or correlation length) of the network, and $\delta$ is the width of the interaction potential. {$\sigma_0$ appears as the stress applied on one link and corresponding to an energy $k_B T$. In other words, $\sigma/\sigma_0$ corresponds to the elastic energy per bond normalized by $k_B T$.} For a network of fractal dimension $D_f$, the characteristic length $\xi$ is linked to the gel concentration $C$ through \cite{Jullien:1987}: $\xi\sim C^{-1/(3-D_f)}$. Assuming that $\delta$ does not significantly depend on $C$, one has $\sigma_0\sim C^{2/(3-D_f)}$. Since our gels are prepared through a strong preshearing step, we may assume that the network fractal dimension is $D_f\simeq 2.3$ as generally observed for shear-induced flocs  \cite{Sonntag:1986,Potanin:1995,Thill:1998}. Such a fractal dimension nicely accounts for the scaling $\sigma_0\sim C^{2.9}$ reported in the present work since $2/(3-D_f)\simeq 2.86$.

More generally, beyond attractive gels and yield stress materials, some viscoelastic fluids such as self-assembled transient networks \cite{Ligoure:2013} also display ``delayed'' dynamics under stress. These materials, composed of supramolecular aggregates (e.g. wormlike micelles, surfactant vesicles, emulsion droplets, etc.) linked together by stickers (most often telechelic polymers), resemble colloidal gels in that their mesoscopic constituents exhibit attractive interactions. However, they do not display any solidlike behavior at rest as the link network is weak and thus temporary \cite{Ligoure:2013}. Interestingly, such materials appear to be ``brittle'' \cite{Tabuteau:2009}: under an imposed shear stress, they exhibit macroscopic fractures which are reversible and occur after a delay time that decreases exponentially for increasing shear stress \cite{Skrzeszewska:2010}. A fiber-bundle like model \cite{Mora:2011} introducing reversible link rupture has been proposed recently to account for fractures in transient networks in a way similar to the model proposed for the activated yielding of colloidal gels \cite{Lindstrom:2012}. One can thus wonder whether there exists some deeper link between the fracture of self-assembled transient networks and the yielding of colloidal gels, and if so, whether these two phenomena can be described in a single generic framework.

\subsection{Open questions}

\subsubsection{Link with standard rheological data.~}The divergence of the creep duration $\tau_c$ as the critical shear stress $\sigma_c$ is approached allows us to clearly define $\sigma_c$ as the yield stress of the material: whatever the shear stress applied below $\sigma_c$ the gel will not start to flow whereas it will eventually get fluidized for $\sigma>\sigma_c$ even if it takes longer and longer as $\sigma$ gets closer to $\sigma_c$. Quantitatively, the fact that $\sigma_c$ lies significantly below the other two estimates $\sigma_{y1}$ and $\sigma_{y2}$ extracted from standard rheological tests is not surprising for such a time-dependent material as CB gels (see Tables~\ref{t.rheol} and \ref{t.fits}). Yet, it is interesting to note that $\sigma_{y2}$ estimated from the flow curve follows roughly the same power law of the gel concentration ($\sim C^{3.5}$) as $\sigma_c$ ($\sim C^{3.7}$). Whether or not this agreement is fortuitous remains an open issue. 

\subsubsection{Interpretation of Andrade creep.~}Based on our observations, an appealing interpretation of the observed Andrade-like creep could be that bulk deformation {is actually fully reversible} (i.e. without any plasticity in the bulk) and that the origin of the power-law creep lies in a two-dimensional process such as stress-induced demixing of the gel into a diphasic system at the inner wall. Indeed, in the different context of pressure solution creep, an analogy with spinodal dewetting has been invoked to explain the Andrade-like creep observed during the indentation of single crystals of sodium chloride in the presence of saturated brine \cite{Dysthe:2002}. Power-law creep was shown to be correlated to the power-law growth of fluid inclusions at the interface. Although very speculative, such a spinodal-like mechanism transposed to our CB gels would imply the growth of colloid-poor (or even pure oil) domains that separate from the bulk colloid-rich material at the inner boundary. In the case of rough boundaries, this growth will eventually lead to total wall slip when the fluid domains extend over the whole height of the Couette cell, while under smooth boundaries the system would slip right from shear start-up. Such a mechanism could also be involved in the creep of carbopol microgels where homogeneous deformation and similar critical-like scaling was reported, although in this case failure at the inner wall was immediately followed by a transient shear-banding regime rather than by a total wall slip regime \cite{Divoux:2011a}.

\subsubsection{Characteristic strains and timescales.~} {Supplemental Fig.~6 shows the data of Fig.~\ref{f.gpt_rough} replotted as a function of the strain $\gamma$. The good collapse of all curves at the end of the initial creep regime (see also inset of Supplemental Fig.~6) suggests that failure at the inner wall at $\tau_c$ can be associated with a characteristic ``yield strain'' $\gamma_c\simeq 0.2$--0.3. The strains $\gamma_f$ corresponding to full fluidization at $\tau_f$ are spread over a very large range $\gamma_f\simeq 200$--3000, which can be expected from the dominance of slip effects in the second and third regimes. Indeed,} it should be kept in mind that the above strains are those indicated by the rheometer. In the presence of dominant wall slip (as is the case here between $\tau_c$ and $\tau_f$), these strains dramatically overestimate the actual strains within the material. {Investigating other concentrations (data not shown) reveals that $\gamma_c$ decreases with $C$ from about 0.5 for the 4\%~w/w CB gel to about 0.2 for $C=10$\%~w/w, while $\gamma_f$ increases from about 100 up to several thousands over the same concentration range and depending on the imposed shear stress.}

In terms of timescales, it remains unclear whether the two times $\tau_c$ and $\tau_f$ revealed in the present work {(or the corresponding strains $\gamma_c$ and $\gamma_f$)} are linked to the two-step yielding observed in attractive glasses \cite{Koumakis:2011} and mentioned in the introduction. In Ref.~ \cite{Koumakis:2011}, the two characteristic yield strains inferred from shear start-up and LAOS experiments fall into the ranges 0.03--0.3 and 1--3 respectively. These much smaller orders of magnitude suggest {\it a priori~}Êdifferent origins and interpretations for the two strains involved in the present experiments and associated with $\tau_c$ and $\tau_f$. However, we recall that experiments in Ref.~ \cite{Koumakis:2011} were performed under imposed shear rate instead of imposed shear stress {and were not complemented by local strain field or velocity field measurements.} It is also worth mentioning that the interpretation of the double yielding process proposed in Ref.~ \cite{Koumakis:2011} has been confirmed only partially by very recent local measurements on Pickering emulsions \cite{Hermes:2013} and certainly deserves more experiments coupled to direct visualizations.

{Finally, the link between the present timescales and those which may be inferred from experiments under controlled shear rate remains to be explored in CB gels. Indeed, several previous studies, e.g. on concentrated suspensions of glass spheres into a polymer matrix \cite{Aral:1994}, on laponite suspensions \cite{Gibaud:2009}, and on carbopol microgels \cite{Divoux:2010,Divoux:2012} have reported a power-law decay of fluidization times with the applied shear rate. In the case of carbopol microgels, a clear link could be made between stress-imposed and strain-imposed fluidization through the steady-state Herschel-Bulkley rheology \cite{Divoux:2011a}. In CB gels, the strong time-dependence of the material is very likely to preclude such a simple link.}

\subsubsection{Role of boundary conditions.~}One last puzzling point is the fact that yielding under rough boundary conditions involves two successive regimes with such different scalings as power law and exponential while the gel is submitted to the same constant stress field. As discussed above, these distinct behaviours hint to different physical processes. We also emphasize that an important difference between the initial creep and the subsequent fluidization regime lies in the boundary conditions: during Andrade-like creep the gel {does not slip against} the walls while it is bounded by viscous lubrication layers during fluidization. Therefore, the role of boundary conditions, such as the surface roughness and the interactions between the wall and the colloidal gel, raises critical open questions. For instance, when smooth walls are used, we can no longer define $\tau_c$ properly and the gel may show slippage well below the yield stress of the bulk material. Clearly, much more work is required to fully understand such an effect of boundary conditions. We believe that a major step forward will be performed through a systematic investigation of yielding under controlled roughness and chemical properties of the cell walls based on microscopic experiments close to the walls. On the theoretical side, recent approaches based on phenomenological models such as the Soft Glassy Rheology or fluidity models or based on constitutive equations such as the Rolie-Poly model have provided promising predictions for yielding timescales that show power-law dependence and are associated with shear-banded flows \cite{Moorcroft:2013}. However, since failure and slippage at the walls turn out to be central to the yielding process of colloidal gels, future theoretical developments still need to include the presence of bounding walls and the specific interactions of the material with the surface in order to account for all the complexity of experimental situations.

\section{Conclusion}

We have reported an extensive set of experiments coupling rheology and velocimetry during creep and yielding of attractive colloidal gels. Our results reveal that under rough boundary conditions yielding proceeds in two successive steps. (i)~The gel first undergoes a creep regime which is fully similar to that reported in some crystalline solids as well as other soft amorphous solids and characterized by a succession of Andrade creep (primary creep), quasi stationary regime (secondary creep), and acceleration of the strain rate (tertiary creep) prior to rupture, here localized at the inner wall of our Couette geometry. During Andrade creep, the bulk strain field appears to remain homogeneous down to the micron-scale. (ii)~The gel then slips totally at both walls and is progressively broken down into smaller and smaller pieces through a strongly heterogeneous flow until full fluidization is reached. The two timescales associated to this yielding scenario follow very different scalings with the applied stress. While the creep duration is governed by a critical-like behaviour, the --generally much larger-- fluidization time follows an exponential decay. These scalings suggest that two different physical mechanisms are successively at play in each step, (i)~either local yielding in the bulk above some critical yield strain as invoked in fiber-bundle models or two-dimensional stress-induced demixing close to the walls, followed by (ii)~activated bond-breaking dynamics.

Our study constitutes the first complete description of the so-called ``delayed yielding'' phenomenon at a mesoscopic level. Obviously, it should be pursued through more microscopic investigations especially close to the bounding walls in order to specify the role of boundary conditions in the fluidization process. It should also be extended to other systems, including attractive and repulsive glasses for comparison with previous shear start-up and LAOS experiments and in order to test for universality in the yielding behaviour of soft solids.

The authors wish to thank R.~Buscall, S.~Lindstr{\"o}m, and G.~Ovarlez for insightful discussions. This work was funded by the European Research Council under the European Union's Seventh Framework Programme (FP7/2007-2013) and ERC Grant Agreement No. 258803.

\providecommand*{\mcitethebibliography}{\thebibliography}
\csname @ifundefined\endcsname{endmcitethebibliography}
{\let\endmcitethebibliography\endthebibliography}{}

\clearpage

\addtocounter{figure}{-14}
\addtocounter{section}{-5}

\section*{\Large{Supplemental material}}

\section{Procedure for removing spurious scattering from the cell walls}

In the present work, the Plexiglas cylinders used in our Couette geometry were sand-blasted in order to provide a roughness of about 1~$\mu$m which leads to significant scattering of the incident ultrasonic pulses. This results in spurious fixed echoes in the raw ultrasonic data that get mixed with the echoes backscattered by the moving particles. Such fixed echoes appear as vertical lines in the spatiotemporal diagram of Supplemental Figure~1(a) that shows the successive pressure signals $p(t_{us},t)$ coded in gray levels as a function of the ultrasonic time-of-flight $t_{us}$ (horizontal axis) after a single pulse is sent at time $t$ (vertical axis). Using the cross-correlation algorithm described in Ref.~\cite{Manneville:2004a} on such raw ultrasonic data leads to a dramatic underestimation of the local velocities at the location of these fixed echoes.

In order to remove the undesired fixed echoes before data analysis, we average the ultrasonic signals recorded during the systematic preshear step at  $\gp_p=10^3$~s$^{-1}$ (see Sect.~2.3) and subtract this average to each raw pressure signal recorded subsequently during the actual experiment. The result is shown in Supplemental Figure~1(b). In the averaging process at large shear rate, all contributions from acoustic scatterers within the sheared fluid cancel out and one is left with the static spurious signal. Subtracting this signal to the raw data appears as a very efficient way to remove the spurious contributions of the ultrasonic waves scattered off by the surface roughness of the outer fixed wall.

\begin{figure}[!t]\tt
\centering
\includegraphics[width=0.85\linewidth]{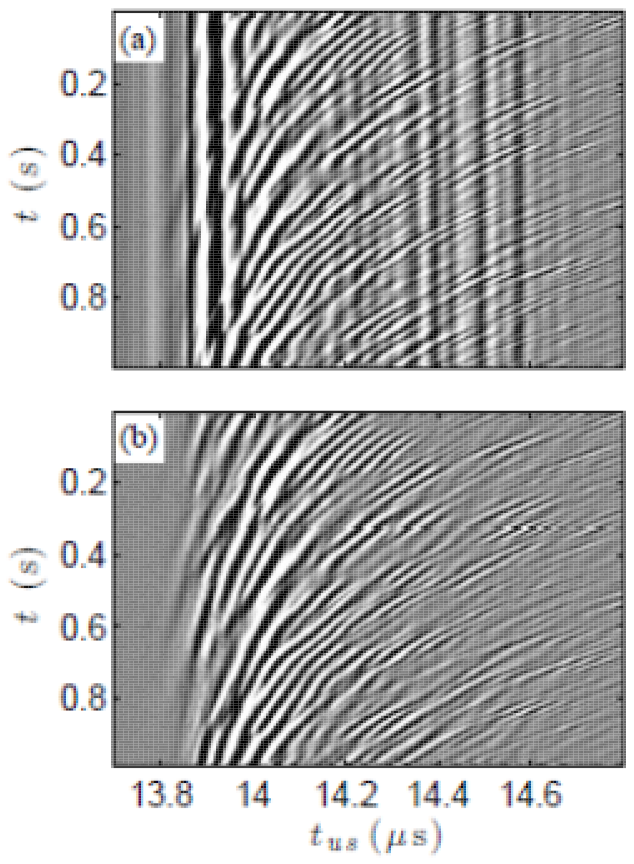}
\caption{Spatiotemporal diagrams of the pressure signal recorded as a function of time $t_{us}$ (horizontal axis) after a single pulse is sent at time $t$ (vertical axis). (a) Raw data. (b) Same data after fixed echoes have been removed following the procedure described in the text. The pressure signal is coded in linear gray levels.}
\label{f.fixed_echoes}
\end{figure}

\newpage
~
\vspace{2cm}

\begin{figure*}[!t]\tt
\centering
\includegraphics[width=0.8\linewidth]{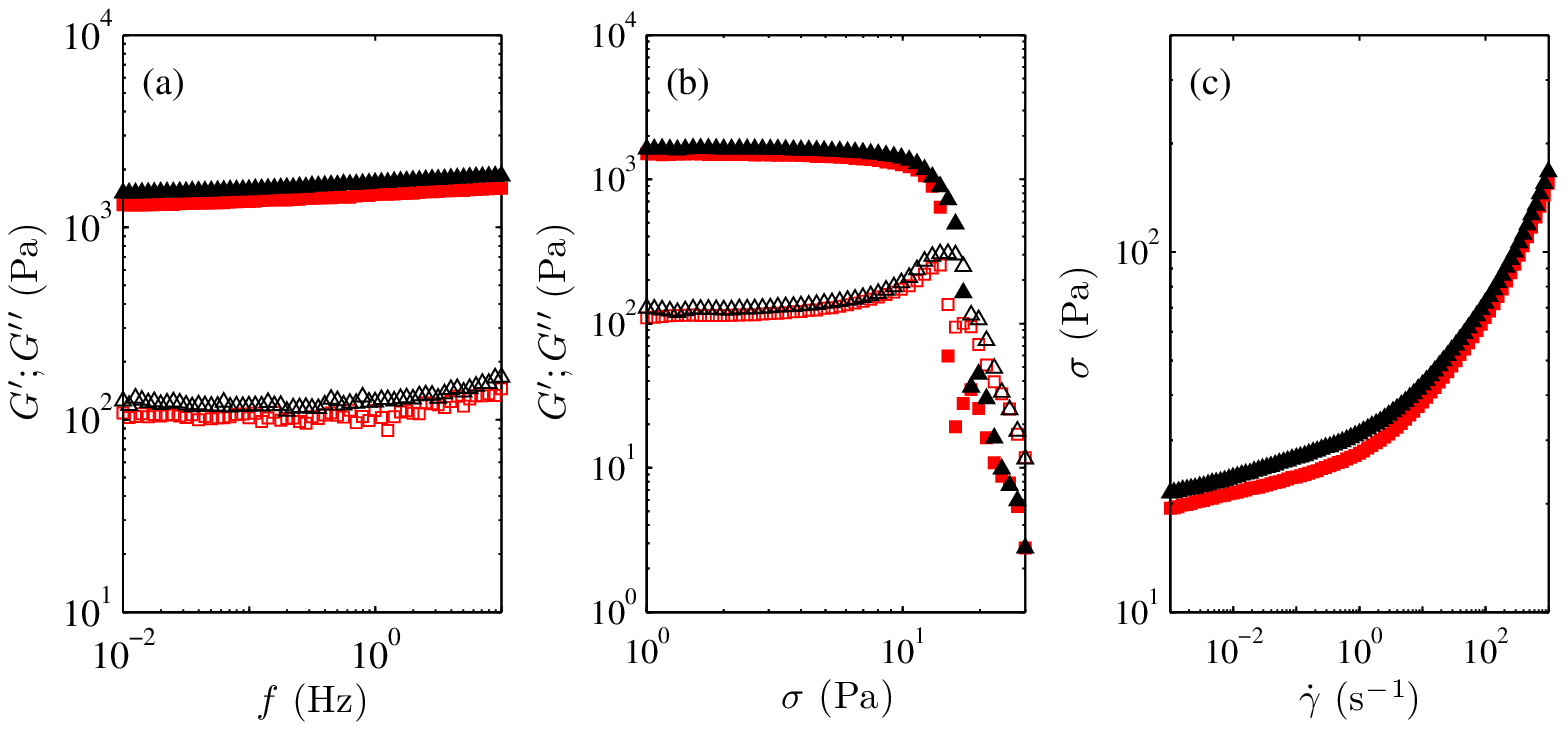}
\caption{Rheological properties of a CB gel at 6\% w/w seeded with  1\% w/w hollow glass microspheres (black triangles) compared to those of the same sample free of seeding microspheres (red squares). Viscoelastic moduli $G'$ (filled symbols) and $G''$ (empty symbols) (a) as a function of frequency $f$ for a stress amplitude of 2~Pa (waiting time of 6 oscillation periods per point) and (b) as a function of stress amplitude $\sigma$ at a frequency of 1~Hz (waiting time of 5~s per point). (c) Flow curves $\sigma$ vs $\gp$ measured by decreasing $\gp$ (waiting time of 1~s per point).}
\label{f.compare_rheol}
\end{figure*}

\section{Influence of adding acoustic contrast agents to carbon black gels}

Supplemental Figure~2 compares the linear and nonlinear rheological properties of 6\% w/w CB gels with and without acoustic contrast agents, namely 1\% w/w hollow glass microspheres of mean diameter 6~$\mu$m (Sphericel, Potters) and density 1.1~g.cm$^3$. Addition of acoustic contrast agents does not significantly affect the mechanical behaviour of CB gels. Quantitatively, we note that the viscoelastic moduli at rest increase by about 10\% upon addition of microspheres [see Supp. Fig.~2(a,b)]. Such an enhancement of the viscoelastic properties is expected and has already been observed in, e.g., carbopol microgels \cite{Divoux:2011b}. Accordingly, the flow curve of a CB gel as well as the yield stress are shifted upwards by about 10\% when adding 1\% w/w hollow glass microspheres to the system [see Supp. Fig.~2(c)]. 

\section{Influence of the preshear protocol on the fluidization time}

\begin{figure}[!b]\tt
\centering
\includegraphics[width=0.85\linewidth]{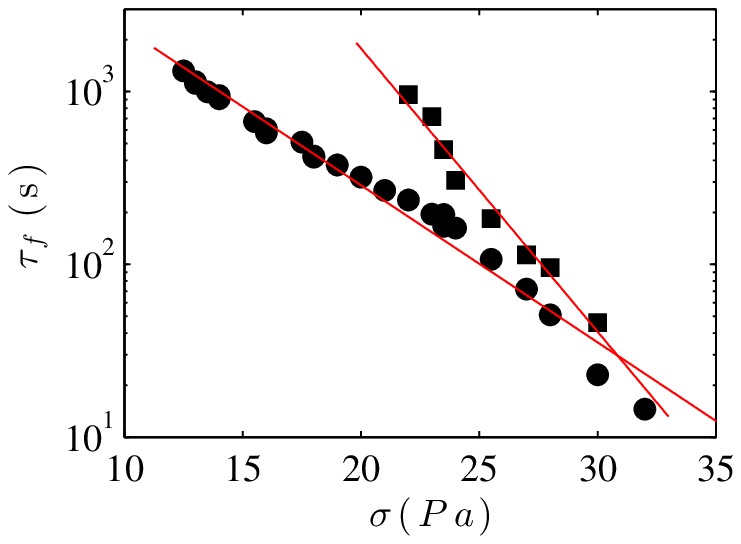}
\caption{Fluidization time $\tau_f$ after two different preshear protocols for a 6\% w/w CB gel under rough boundary conditions: preshear for 20~s at $+1000$~s$^{-1}$ ($\blacksquare$) and at $-1000$~s$^{-1}$ ($\bullet$). The shear stress $\sigma$ is applied in the positive direction once viscoelastic moduli have been measured for 300~s after preshear. Red lines are the best exponential fits $\tau_f=\tau_0\exp(-\sigma/\sigma_0)$.}
\label{f.preshear_direction}
\end{figure}

\begin{figure*}[!t]\tt
\centering
\includegraphics[width=0.85\linewidth]{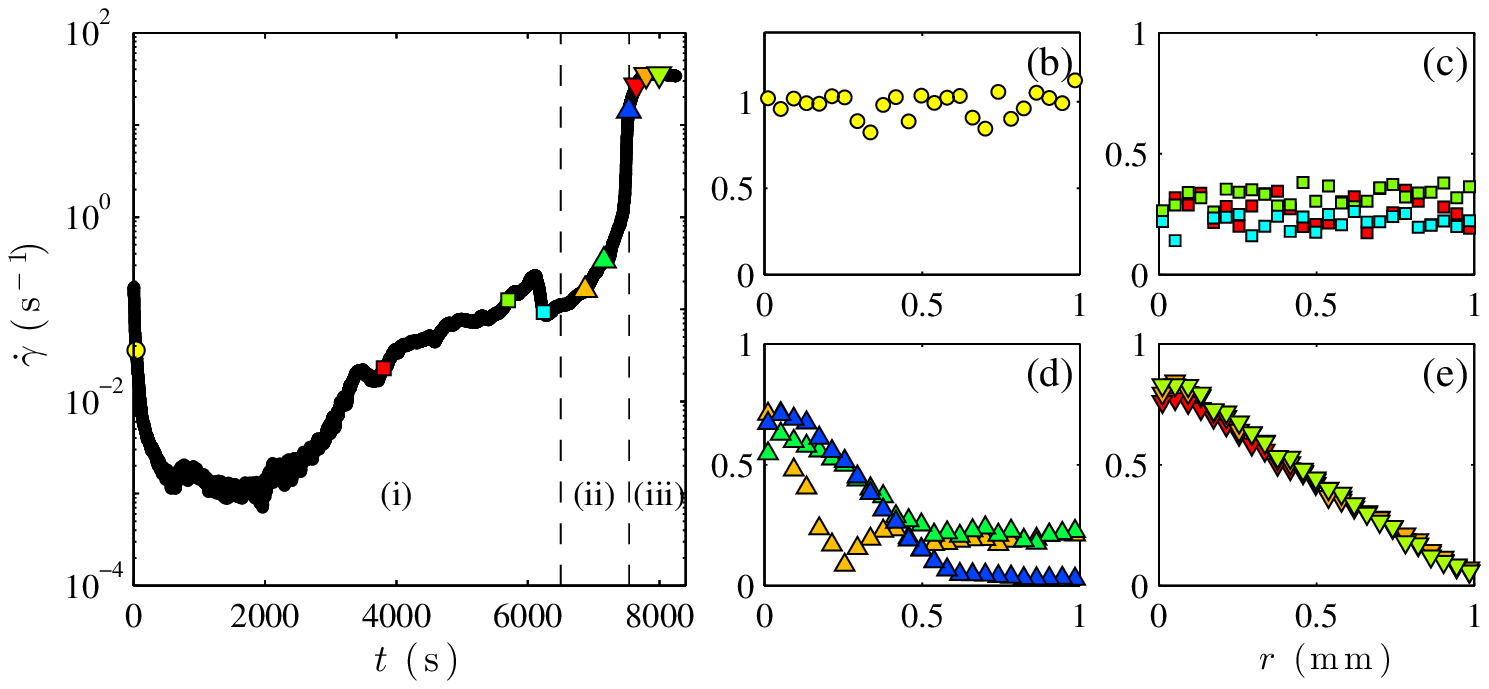}
\caption{Creep experiment in a 6\% w/w CB gel at $\sigma=16$~Pa under smooth boundary conditions. (a) Shear rate response $\gp(t)$. The vertical dashed lines indicate the limits of the three regimes discussed in the text. The coloured symbols show the times at which the velocity profiles in (b)--(e) are recorded. Velocity profiles $v(r,t_0)$, where $r$ is the distance to the rotor, normalized by the current rotor velocity $v_0(t_0)$ at (b)~$t_0=39$~s, (c)~$t_0=3806$, 5700, and 6240~s, (d)~$t_0=6871$, 7157, and 7536~s, and (e)~$t_0=7650$, 7800, and 8000~s.}
\label{f.velocity_smooth}
\end{figure*}

The fluidization time $\tau_f$ was measured as described in the main text after preshearing a 6\% w/w CB gel either at $+1000$~s$^{-1}$ or at $-1000$~s$^{-1}$ for 20~s before viscoelastic moduli at rest are monitored for 300~s and a given stress $\sigma$ is subsequently applied in the positive direction. As shown in Supplemental Figure~3 this leads to significant differences in the yielding phenomenon. In both cases, an exponential behaviour is found for $\tau_f$ vs $\sigma$ but fluidization is much faster when creep and preshear are applied in opposite directions. 

We checked that:\\
{(i)~the preshearing direction does not affect the shape of the subsequent shear rate response $\gp(t)$ (data not shown), which remains similar to the responses shown in Fig.~4 in the main text and, in particular, shows three well-defined regimes.}\\
 (ii)~reversing both preshear and creep directions does not affect $\tau_f$ so that the difference may not be attributed to an artifact due to our rheometer or geometry,\\
(iii)~in the case of successive preshears with different directions such as in the protocol used in the main text ($+1000$~s$^{-1}$ followed by $-1000$~s$^{-1}$), the fluidization time is only affected by the last preshear step.\\
This clearly shows that, even though preshear successfully erases previous sample history, the resulting gel microstructure is sensitive to preshear.

The influence of preshear was investigated by Osuji {\it et al.} \cite{Osuji:2008b} in CB gels in tetradecane at  2--8\% w/w. A power-law dependence of the elastic modulus with the shear stress applied during preshear was reported together with a slow decrease of the residual ``internal stress'', i.e. the shear stress measured after flow cessation, $\sigma_i(t)\sim t^{-0.1}$. These findings were interpreted based on a simple model for the cluster size reached after preshearing at a stress $\sigma_p$ and on an unusually fast build-up of the network structure after cessation of shear in which internal stresses act opposite to the preshear direction. 

Internal stresses may partly explain our results. Indeed, if stress is applied in the direction opposite to preshear, internal stress adds up to the applied stress, thus facilitating yielding and leading to a faster fluidization process. Yet Supplemental Figure~3 shows that the effect of preshear is not simply an effective change of $\sigma$ by a constant $\pm\sigma_i$ depending on the preshear direction since in this case the two curves $\tau_f$ vs $\sigma$ would only be translated by a constant amount. Moreover, if the differences in fluidization times were to be explained solely by internal stresses, then one would expect that for very long fluidization times (i.e. for small $\sigma$), the slow relaxation of internal stresses leads to smaller discrepancies in $\tau_f$. This is not observed in our data. Rather, fluidization times become similar for large values of $\sigma$ and both parameters $\sigma_0$ and $\tau_0$ in the exponential fits depend on the preshear direction. We find $\sigma_0=2.6$~Pa and $\tau_0=3.4\,10^6$~s for a preshear in the positive direction and $\sigma_0=4.8$~Pa and $\tau_0=1.9\,10^4$~s for the opposite direction (see red lines in Supp. Figure~3). This suggests that the anisotropy of the gel structure induced by preshearing plays an important role in the delayed fluidization under creep. Such an anisotropy is not accounted for in the model of Ref.~\cite{Lindstrom:2012}.

\section{Velocity profiles under smooth boundary conditions}

Supplemental Figure~4 reports velocity profiles recorded during a creep experiment performed under smooth boundary conditions on a 6\% w/w CB gel together with the corresponding evolution of the shear rate $\gp(t)$ [see Supp. Figure~4(a)]. Total slippage at the fixed outer wall is observed as soon as shear is applied at $t=0$ [see Supp. Fig.~4(b)]. Although velocities for $200\lesssim t\lesssim 3000$~s are too small to allow for reliable measurements, the flow is most likely pluglike throughout the creeping flow regime (i) with slip velocities increasing at the rotor and decreasing at the stator. Indeed, once the shear rate has raised above roughly $10^{-2}$~s$^{-1}$ allowing velocities to be accurately estimated, velocities show a flat profile with almost total slippage at the rotating inner wall [see Supp. Fig.~4(c)]. After a small bump in $\gp(t)$ which is characteristic of the shear rate response in a smooth cell (here at $t\simeq 6000$~s, see also Fig.~10 in the main text), highly fluctuating shear-banded velocity profiles are recorded [regime (ii), see Supp. Fig.~4(d)]. Steady homogeneous velocity profiles are recovered after the inflection point in $\gp(t)$, with about 10\% of residual wall slip at the rotor [regime (iii), see Supp. Fig.~4(e)].

\section{Evolution of the characteristic shear rate $\gp^\star$ after failure at the inner wall}

The characteristic shear rate $\gp^\star$ after failure at the inner wall at $t=\tau_c$ under rough boundary conditions is shown in Supplemental Figure~5 as a function of the applied shear stress $\sigma$ for four gel concentrations $C$. $\gp^\star$ is seen to increase fairly linearly with the applied stress $\sigma$ and, on average, to increase with the gel concentration $C$.

\begin{figure}[!h]\tt
\centering
\includegraphics[width=0.85\linewidth]{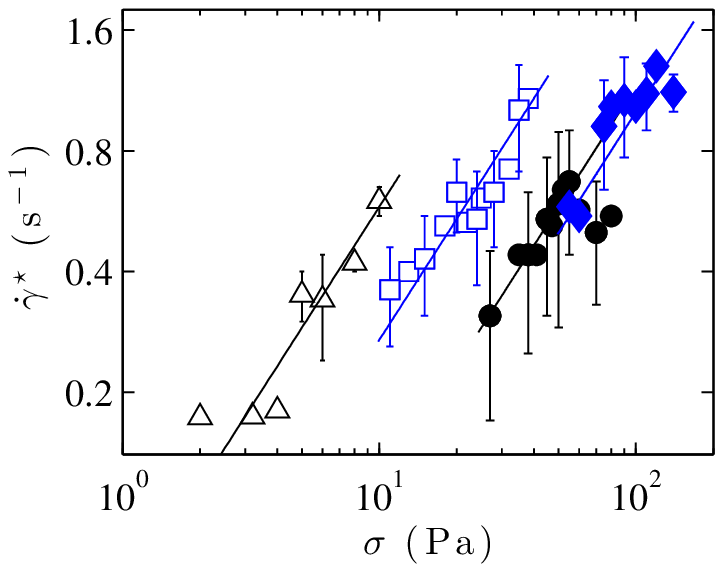}
\caption{Shear rate $\gp^\star$ after failure at the inner wall at $t=\tau_c$ as a function of the applied shear stress $\sigma$ for CB gels of concentration $C=4$ ($\triangle$), 6 ($\square$), 8 ($\bullet$), and 10\%~w/w~($\blacklozenge$). Solid lines correspond to linear behaviours $\gp^\star\propto\sigma$. Error bars show the variations of $\gp(t)$ over the shear rate plateau for $\tau_c< t < \tau_f$.}
\label{f.gpstar}
\end{figure}

\section{Shear rate response as a function of strain}

{Supplemental Fig.~6 shows the data of Fig.~\ref{f.gpt_rough} replotted as a function of the strain $\gamma$. The good collapse of all curves at the end of the initial creep regime (see also inset of Supp. Fig.~6) suggests that failure at the inner wall at $\tau_c$ can be associated with a characteristic ``yield strain'' $\gamma_c\simeq 0.2$--0.3. The strains $\gamma_f$ corresponding to full fluidization at $\tau_f$ are spread over a very large range $\gamma_f\simeq 200$--3000.}

\begin{figure*}[!t]\tt
\centering
\includegraphics[width=0.8\linewidth]{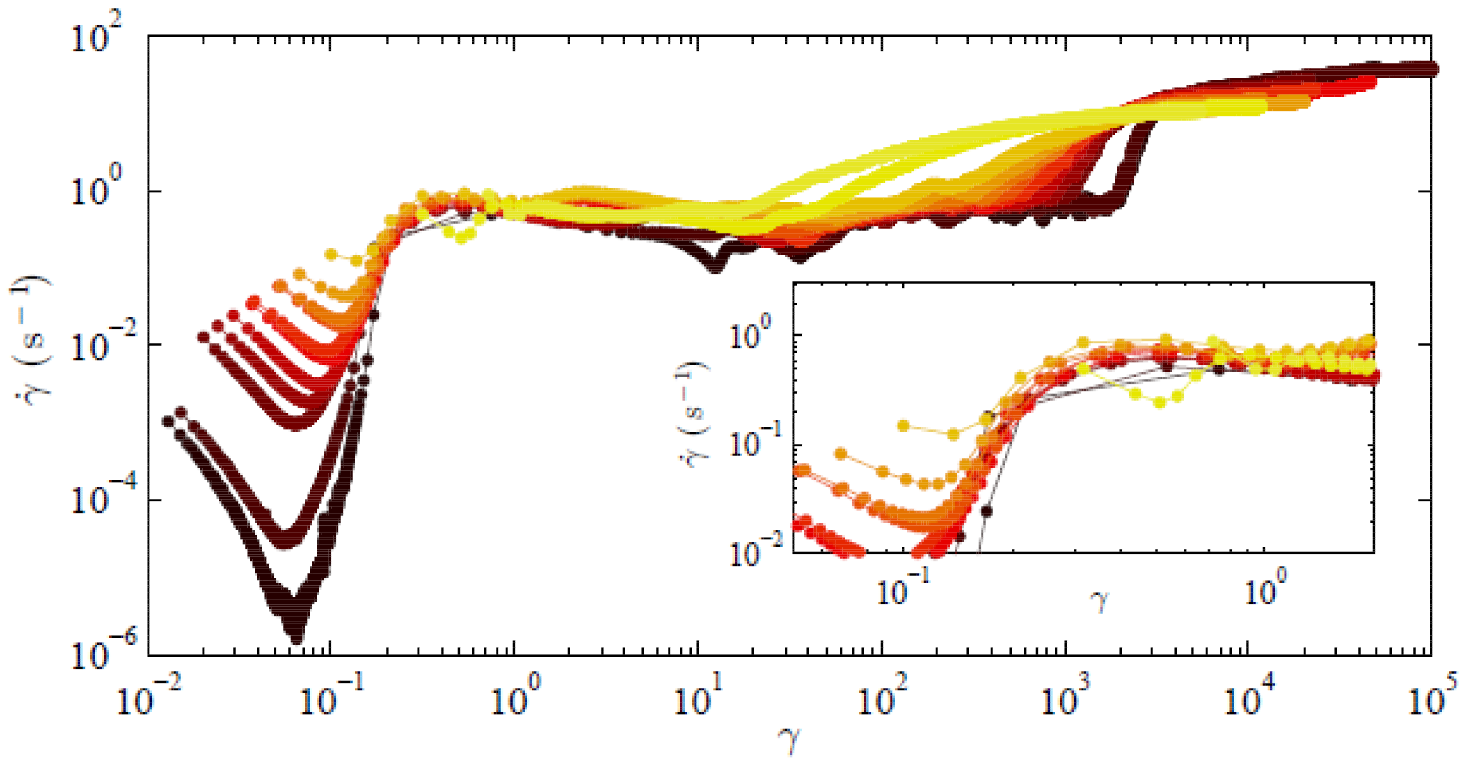}
\caption{Creep experiments in an 8\%~w/w CB gel under rough boundary conditions. Shear rate responses $\gp$ as a function of the strain $\gamma$ for different shear stresses $\sigma$ applied at time $t=0$: from right to left, $\sigma=24$, 27, 35, 38, 41, 45, 47, 50, 52, 55, 60, 70, and 80~Pa. Inset: enlargement over the end of the initial creep regime.}
\label{f.gpt_rough}
\end{figure*}

\end{document}